\documentclass[conference]{IEEEtran}
\IEEEoverridecommandlockouts

\usepackage{cite}
\usepackage{amsmath,amssymb,amsfonts}
\usepackage{algorithmic}
\usepackage{graphicx}
\usepackage{textcomp}
\usepackage{multirow}
\usepackage{xspace}
\usepackage{tikz}
\usepackage{tcolorbox}
\usepackage{tabularx}
\usepackage{tabularx}
\usepackage{booktabs}
\usepackage{subfig, xcolor}
\usepackage{xurl}
\usepackage{algorithmic}
\usepackage{multirow}
\usepackage{makecell, booktabs}

\usepackage{url}
\usepackage{enumitem}
\usepackage{flushend}
\usepackage[bookmarks=true,breaklinks=true,letterpaper=true,colorlinks,citecolor=blue,linkcolor=blue,urlcolor=blue]{hyperref}

\makeatletter
\renewcommand{\footnoterule}{%
  \kern -3pt
  \hrule width 0.4\columnwidth
  \kern 2.6pt
}
\makeatother

\usepackage[linesnumbered,ruled,vlined]{algorithm2e}
\def\BibTeX{{\rm B\kern-.05em{\sc i\kern-.025em b}\kern-.08em
    T\kern-.1667em\lower.7ex\hbox{E}\kern-.125emX}}
\newcommand\copyrighttext{%
	\footnotesize This is the preprint accepted for publication in the 26th IEEE International Symposium
    on Cluster, Cloud, and Internet Computing (CCGrid 2026), Sydney, Australia, 18–21 May 2026. This version is released under a CC-BY license in accordance with the Horizon Europe programme requirements.}
\newcommand\copyrightnotice{%
	\begin{tikzpicture}[remember picture,overlay]
		\node[anchor=north,yshift=-10pt] at (current page.north) {\fbox{\parbox{\dimexpr\textwidth-\fboxsep-\fboxrule\relax}{\copyrighttext}}};
	\end{tikzpicture}%
}

\begin{document}
\font\titlefont=cmr12 at 20pt
 
\title{ 
\titlefont
Green or Fast? Learning to Balance Cold Starts and Idle Carbon in Serverless Computing
}

\author{
\IEEEauthorblockN{Bowen Sun\textsuperscript{*}}
\IEEEauthorblockA{\textit{Department of Computer Science} \\
\textit{William \& Mary}\\
Williamsburg, VA, USA \\
bsun02@wm.edu}
\and
\IEEEauthorblockN{Christos D. Antonopoulos}
\IEEEauthorblockA{\textit{Department of Computer Science} \\
\textit{University of Thessaly}\\
Volos, Greece \\
cda@uth.gr}
\and
\IEEEauthorblockN{Evgenia Smirni}
\IEEEauthorblockA{\textit{Department of Computer Science} \\
\textit{William \& Mary}\\
Williamsburg, VA, USA \\
exsmir@wm.edu}
\and[\hfill\mbox{}\par\mbox{}\hfill]
\IEEEauthorblockN{Bin Ren}
\IEEEauthorblockA{\textit{Department of Computer Science} \\
\textit{William \& Mary}\\
Williamsburg, VA, USA \\
bren@wm.edu}
\and
\IEEEauthorblockN{Nikolaos Bellas}
\IEEEauthorblockA{\textit{Department of Computer Science} \\
\textit{University of Thessaly}\\
Volos, Greece \\
nbellas@uth.gr}
\and
\IEEEauthorblockN{Spyros Lalis}
\IEEEauthorblockA{\textit{Department of Computer Science} \\
\textit{University of Thessaly}\\
Volos, Greece \\
lalis@uth.gr}
\thanks{\textsuperscript{*}Ph.D. candidate at William and Mary. This work was completed while the author was a researcher at University of Thessaly, Greece.}
}

\newcommand{\framename}{\text{LACE-RL}\xspace}
\newcommand{\framenames}{\text{LACE-RL's}\xspace}
\newcommand*\circled[1]{\tikz[baseline=(char.base)]{\node[shape=circle,draw,fill=black,text=white,font=\bf,minimum size=3mm,inner sep=0pt] (char) {#1};}}

\maketitle
\copyrightnotice
\thispagestyle{plain}
\pagestyle{plain}

\begin{abstract}
Serverless computing simplifies cloud deployment but introduces new challenges in managing service latency and carbon emissions. Reducing cold-start latency requires retaining warm function instances, while minimizing carbon emissions favors reclaiming idle resources. This balance is further complicated by time-varying grid carbon intensity and varying workload patterns, under which static keep-alive policies are inefficient.
We present \framename, a latency-aware and carbon-efficient management framework that formulates serverless pod retention as a sequential decision problem. \framename uses deep reinforcement learning to dynamically tune keep-alive durations, jointly modeling cold-start probability, function-specific latency costs, and real-time carbon intensity.
Using the Huawei Public Cloud Trace, we show that \framename reduces cold starts by $51.69\%$ and idle keep-alive carbon emissions by $77.08\%$ compared to Huawei’s static policy, while achieving better latency–carbon trade-offs than state-of-the-art heuristic and single-objective baselines, approaching Oracle performance.
\end{abstract} 


\section{Introduction}
\label{sec:intro}

Serverless computing has become a widely adopted paradigm for cloud applications, offering elastic scaling and simplified deployment by abstracting infrastructure provisioning and management~\cite{aws_lambda, azure_functions}. Once viewed as a niche technology, serverless is now used in production by over 50\% of organizations~\cite{datadog_state_of_serverless}. As serverless platforms scale, however, they must increasingly reconcile service performance with energy efficiency and environmental impact. In particular, cold-start latency~\cite{liu2023faaslight} and the carbon footprint of idle {\textit{pods}}\footnote{A pod represents the basic container instance that executes a serverless function; we use it synonymously with a function container or instance.} that host functions~\cite{joosen2025serverless, stojkovic2024ecofaas} have emerged as closely related challenges shaped by pod keep-alive decisions that determine whether pods are retained or reclaimed after function execution.

Cold starts occur when a function is invoked without a warm pod, requiring initialization of both the execution environment and runtime~\cite{zhang2021faster, cvetkovic2024dirigent}. This process can introduce hundreds of milliseconds to several seconds of additional latency, potentially violating service-level objectives (SLOs) in latency-sensitive applications~\cite{bhasi2024paldia, aslani2025machine, qi2024casa}. Our analysis of the recently released Huawei Cloud Trace~\cite{joosen2025serverless}, the most up-to-date public dataset on serverless cold starts, reveals substantial diversity in both pod reuse patterns and cold-start latencies across functions. This variability limits the effectiveness of static keep-alive policies~\cite{shahrad2020serverless, joosen2025serverless} and motivates decision-making that accounts for function-specific behavior.

Reducing cold-start latency typically requires maintaining warm idle pods. However, both our empirical characterization and recent studies~\cite{roy2024hidden} show that aggressive keep-alive strategies can incur a substantial “hidden” carbon cost, in some cases exceeding the emissions associated with function execution itself. Moreover, the carbon footprint of keeping idle pods warm varies over time with changes in grid carbon intensity~\cite{electricitymaps}. Despite this temporal variability, production platforms continue to rely on static keep-alive policies, such as fixed timeouts used in Huawei Cloud~\cite{joosen2025serverless}, or simple heuristic pre-warming. Such approaches are ill-suited to environments with diverse function behavior, non-stationary workloads, and time-varying carbon intensity, where effective keep-alive decisions require adapting to both workload dynamics and external conditions (the time-varying carbon footprint).

The above observations motivate an adaptive control framework for keep-alive management, which makes per-function invocation decisions by considering pod reuse potential, cold-start latency, and time-varying carbon intensity. Keeping a pod warm reduces cold-start latency but increases idle energy use and carbon emissions. As a result, keep-alive decisions form a sequential optimization problem under uncertainty, with long-term latency–carbon trade-offs: a decision made now determines pod availability (warm vs.\ cold) for future invocations as well as idle energy consumption and associated carbon emissions. Effective keep-alive control must operate beneath cluster-level autoscaling, enabling fine-grained pod management and more aggressive downscaling. While prior work examines cold-start mitigation~\cite{vahidinia2022mitigating, ebrahimi2024cold}, pre-warming, or carbon-aware management~\cite{gsteiger2024caribou, qi2024casa} in isolation~\cite{shafiei2022serverless, patros2021toward}, \emph{no existing system optimizes keep-alive decisions at function-instance granularity under time-varying carbon intensity}.

We address this gap with \framename, a \underline{L}atency-\underline{A}ware and \underline{C}arbon-\underline{E}fficient serverless management framework based on \underline{R}einforcement \underline{L}earning. \framename learns per-function keep-alive policies that balance latency and carbon objectives while adapting to diverse workloads and changing carbon conditions. Our key contributions are summarized below.
\begin{itemize}[leftmargin=10pt]
\item \textit{Formalization of the latency--carbon trade-off.} We formalize serverless pod retention as a per-pod, multi-objective sequential decision problem that captures cold-start probability, function-specific latency penalties, and time-varying grid carbon intensity.
\item \textit{RL-based adaptive keep-alive control.} We design a reinforcement learning framework that learns per-pod keep-alive policies from workload dynamics and environmental signals. Unlike static or heuristic approaches, \framename adapts to varying cold-start behaviors and supports configurable latency--carbon trade-offs.

\item \textit{Cold-start and energy profiling.} We profile a controlled FunctionBench~\cite{kim2019functionbench} deployment using Kepler~\cite{kepler2025} to calibrate phase-level energy accounting, characterize diverse cold-start costs, and quantify non-trivial energy consumption during keep-alive periods, grounding both our simulator and optimization objectives in real-machine measurements.

\item \textit{Comprehensive evaluation.} We evaluate \framename using a trace-driven simulator based on the Huawei Cloud Trace~\cite{joosen2025serverless} and calibrated with empirical measurements. \framename reduces cold starts by $51.69\%$ and idle keep-alive carbon emissions by $77.08\%$ relative to Huawei’s static policy~\cite{joosen2025serverless}, and achieves superior latency--carbon trade-offs compared to single-objective baselines and state-of-the-art approaches~\cite{jiang2024ecolife}, while approaching Oracle performance.
\end{itemize}

\section{Energy Modeling and Challenges}
\label{sec:background}

\begin{table*}[t]
  \caption{Summary of Huawei Dataset Components Used in the \framename Framework}
  \label{tab:huawei-dataset}
  \centering
  \begin{tabular}{>{\raggedright\arraybackslash}p{3cm} >{\raggedright\arraybackslash}p{6.5cm} >{\raggedright\arraybackslash}p{5.5cm}}
    \toprule
    Dataset Source & Description & Usage in \framename Framework \\
    \midrule
    Cold Start Logs & 
    31-day cold start traces with breakdowns of \texttt{pod allocation cost}, \texttt{code deploy cost}. & 
    Used to construct function-specific cold start latency ($L_{cs}$) by runtime \\
    
    Request-Level Logs & 
    Per-invocation metadata including \texttt{timestamp}, \texttt{execution time}, \texttt{podID}, \texttt{CPU}, and \texttt{memory} request & 
    Used to simulate invocation traces, calculate pod reuse intervals, and model pod reuse probabilities \\
    
    Runtime and Trigger Metadata & 
    Static table mapping each function with its runtime language and trigger type & 
    Used to determine runtime-sensitive latency and filter function categories \\
    \bottomrule
  \end{tabular}
\end{table*}

We analyze a public serverless trace from Huawei Cloud~\cite{joosen2025serverless} that includes fine-grained invocation patterns, pod reuse behavior, and diverse cold-start latencies. Then, we  introduce the energy model for estimating carbon emissions in \framename. The proposed energy model integrates pod-level resource utilization and temporal carbon intensity, enabling a systematic exploration of performance–sustainability trade-offs.

\subsection{Huawei Cloud Dataset}

We analyze the Huawei Cloud Trace~\cite{joosen2025serverless}, focusing on day 30, which includes more than 300 million detailed request-level records~\cite{huawei2025}. We select this dataset over alternatives like Azure~\cite{shahrad2020serverless} or Alibaba~\cite{weng2022mlaas} due to two key advantages: (i) unique insights into cold start latency diversity and (ii) timeliness, ensuring that it reflects modern workloads.

Unlike other public datasets, this trace includes per-invocation timestamps and pod identifiers. These features enable us to infer pod reuse patterns and strictly distinguish between cold and warm starts.  Moreover, it contains cold start latency components and detailed resource configurations. This granularity supports the pod-centric modeling required for the adaptive strategy of \framename. Table~\ref{tab:huawei-dataset} summarizes the structure of the dataset used.

\subsection{Energy Model: Accounting for Serverless Carbon Footprint}
\label{subsec:energymodel}
We estimate per-invocation carbon emissions using runtime hardware activity and grid carbon intensity, based on the energy and carbon accounting model proposed in~\cite{roy2024hidden, jiang2024ecolife}. We summarize the model here for completeness and to aid readability.

We focus exclusively on \textit{operational energy} and exclude embodied carbon (e.g., manufacturing) as it is invariant to runtime retention strategies.
We also assume a \textit{homogeneous hardware environment} to isolate management effects from hardware heterogeneity.
Our model divides energy consumption into two phases: pod \textit{execution} and pod \textit{keep-alive} to reduce cold starts.

\textbf{Function Execution Energy.} Per-invocation execution energy for function $f$ is estimated as:
\begin{equation}
    E_{\text {exec }}(f)=(J_{D R A M}^{M B} \cdot m e m_f+J_{C P U}^{\text {core }} \cdot c p u_f) \cdot T_{\text {exec }}(f),
\end{equation}
where $mem_f$ and $cpu_f$ denote memory (MB) and CPU (cores) allocation, $T_{exec}(f)$ is execution time, and $J$ represents power per MB of memory and per CPU core respectively.
The carbon emitted is $C_{exec}(f,t)=E_{exec}(f) \cdot CI(t)$, where $CI(t)$ is the carbon intensity at time window $t$.
We use real-time carbon intensity data from Electricity Maps~\cite{electricitymaps}, a real-time, data-driven platform that reports per region carbon intensity (grams of $CO_2$ equivalent emissions per kWh of electric energy) based on the variations of fuel mix used and availability of renewable sources.
We assume the carbon intensity $CI(t)$ remains constant during a short execution window. 

\textbf{Keep-alive Energy.} Retaining warm pods for function $f$ mitigates cold starts but incurs idle energy:
\begin{equation}
    E_{\text {idle }}(f)=(J_{D R A M}^{M B} \cdot \text { mem }_f+J_{\text {CPU }}^{\text {core }} \cdot c p u_f) \cdot T_{\text {idle }}(f),
\end{equation}
where $T_{idle}(f)$ is the retention duration. Since idle pods consume less power than active ones, we apply a scaling factor $\lambda_{idle}$~\cite{roy2024hidden}:
\begin{equation}
    E_{\text {idle }}^{scaled}(f)= \lambda_{idle} \cdot E_{\text {idle }}(f)
\end{equation}
We set $\lambda_{idle}=0.2$ based on prior studies~\cite{jiang2024ecolife, stojkovic2024ecofaas}. 
The choice of $\lambda_{idle}=0.2$ is further justified in Sec.~\ref{subsec:exp-setup} via actual experiments on a real system.
The resulting carbon footprint is $C_{idle}(f,t)=E_{\text {idle }}^{scaled}(f) \cdot CI(t)$.

\textbf{Cold-start Energy.} Starting pods to serve an invocation of function $f$ incurs cold-start energy:
\begin{equation}
    E_{\text {cold }}(f)=P_{cold}(f) \cdot T_{\text {cold }}(f),
\end{equation}
where $T_{cold}(f)$ is the latency of the cold start and $P_{cold}$ is the power consumed during cold start. Although $P_{cold}$ is not strictly equal to the execution power, the empirical characterization discussed in Sec.~\ref{subsec:exp-setup} shows that this term is dominated by the value of $T_{cold}(f)$. The resulting footprint is $C_{cold}(f,t)=E_{cold}(f) \cdot CI(t)$.

\textbf{Aggregated Carbon Accounting.} The total carbon footprint is $C_{total}(f,t)=C_{exec}(f,t)+C_{idle}(f,t)+C_{cold}(f,t)$.

\subsection{Challenges in Serverless Management}
\label{subsec:motivation}
We identify three key challenges that complicate serverless management by analyzing patterns from the Huawei Cloud Trace~\cite{joosen2025serverless} and Electricity Maps~\cite{electricitymaps}.

\begin{figure}[t]
    \centering
    \subfloat[CDF of average reuse interval per pod.]{\label{subfig:interarrcdf}\includegraphics[width=0.48\columnwidth]{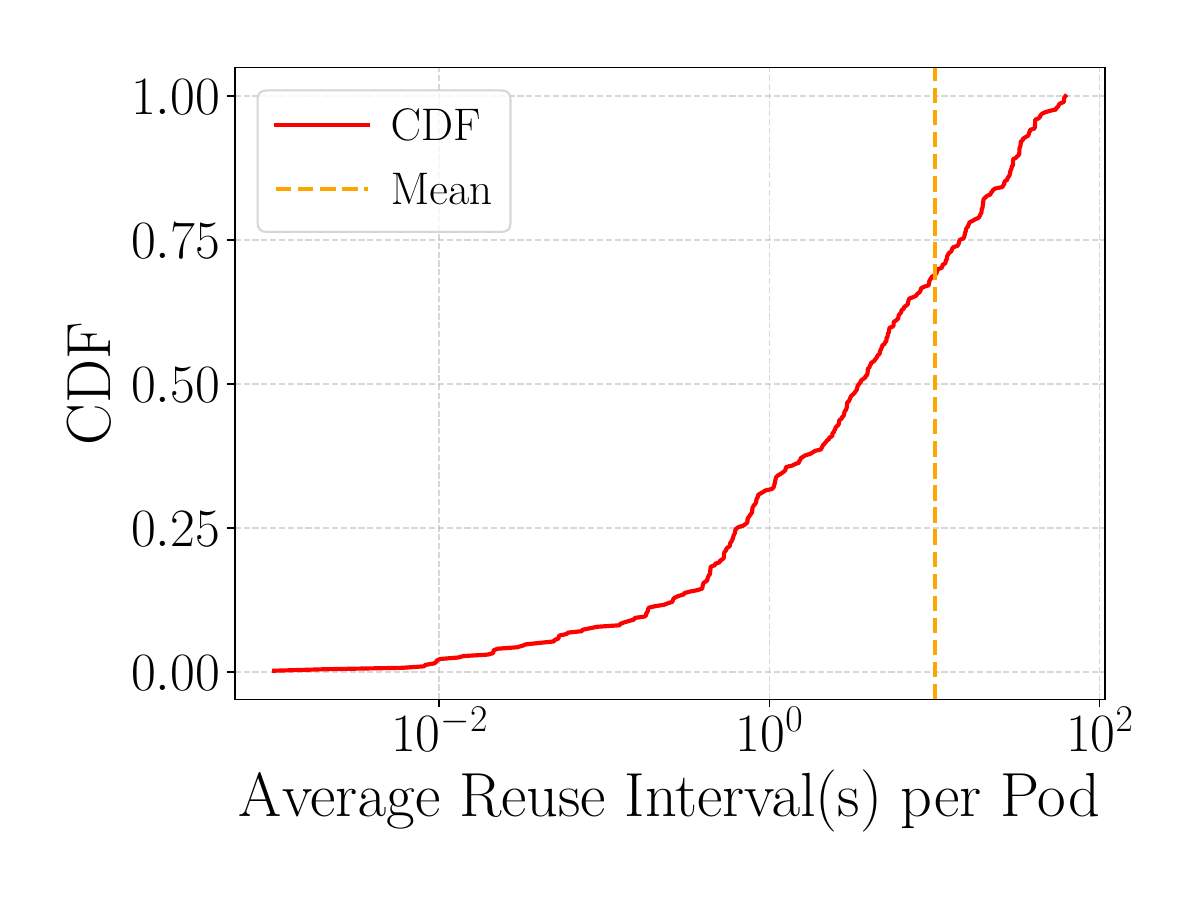}}
    \hspace{0.02\columnwidth}
    \subfloat[{Cold start latency CDF. The gray area highlights distribution ``tail".}]{\label{subfig:cslatencycdf}\includegraphics[width=0.48\columnwidth]{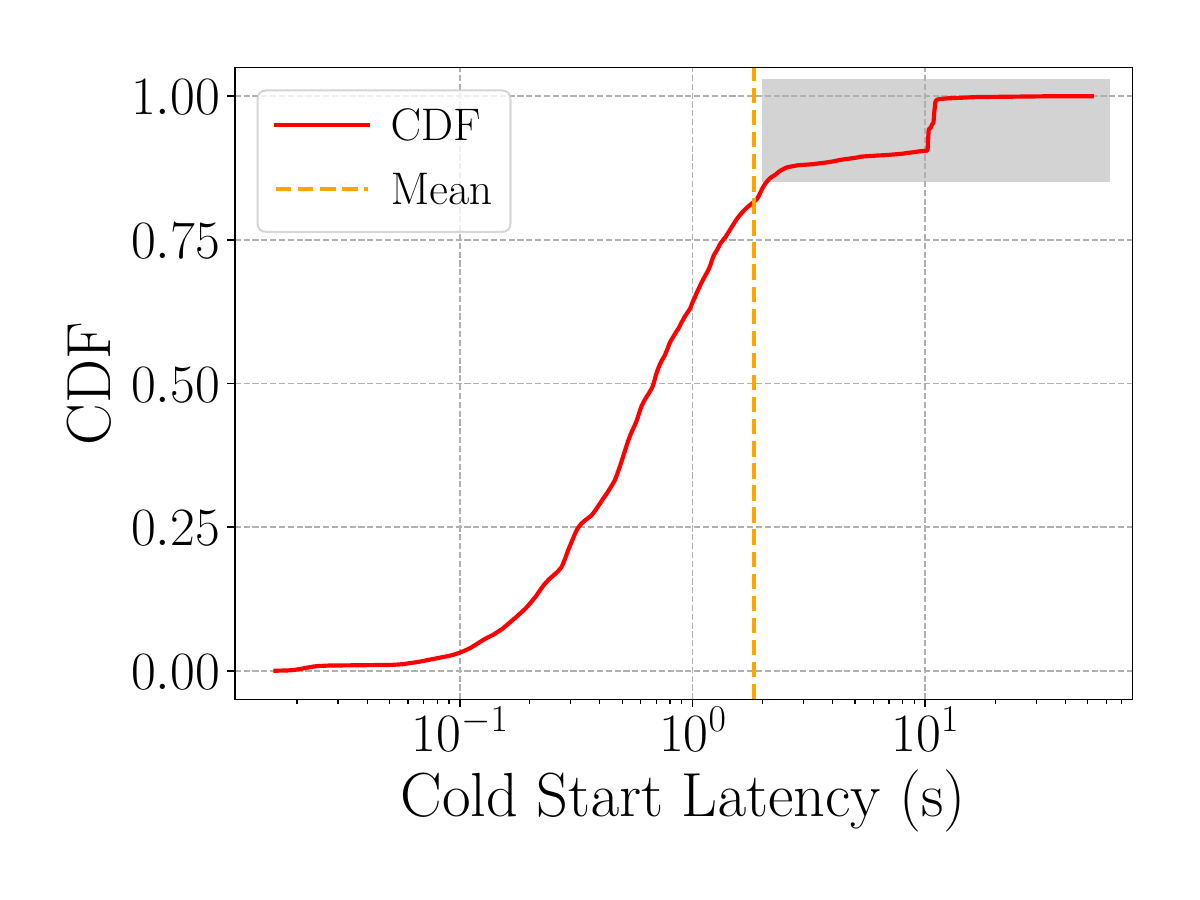}}
    \caption{Characterization of the Huawei Cloud Trace~\cite{joosen2025serverless}.}
    \label{fig:interarr-cslatency}
\end{figure}

\textbf{Challenge I: Irregular Pod Reuse and Cold Start Dynamics.}
We observe high variability in pod invocation patterns. Fig. \ref{subfig:interarrcdf} shows the cumulative distribution function (CDF) of the average time across successive pod invocations. The reuse intervals range from milliseconds to hundreds of seconds, confirming that a single keep-alive timeout cannot perform well across all functions. 
Similarly, Fig. \ref{subfig:cslatencycdf} illustrates that cold start latencies vary from less than 0.1s to more than 10s, depending on the runtime and deployment context. This variance invalidates the fixed latency assumptions made in other work~\cite{jiang2024ecolife, lee2024spes}. Capturing the distribution tail is critical, motivating adaptive management strategies rather than static ones.

\begin{figure}[t]
    \centering
    \includegraphics[width=\columnwidth]{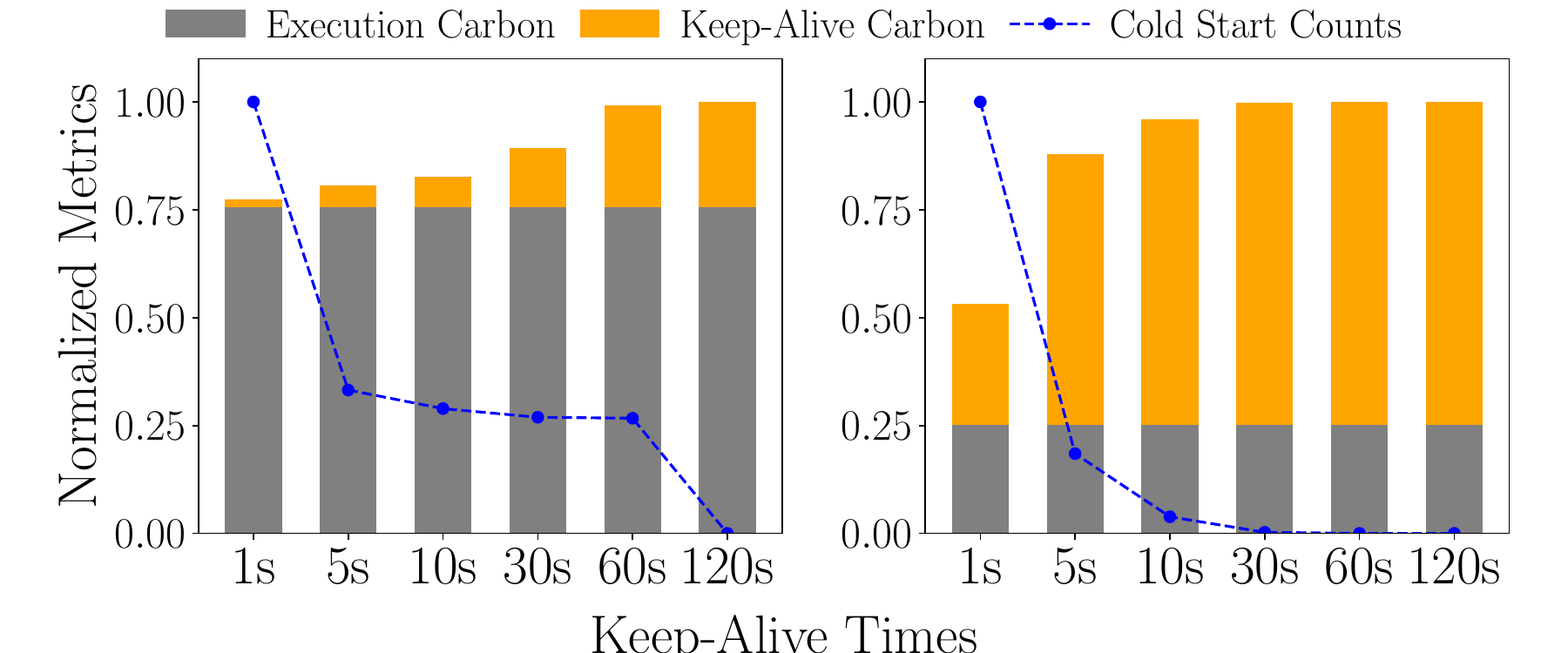}
    \caption{
      Impact of keep-alive timeout for two different but representative functions in the Huawei dataset. In both cases, longer timeouts reduce cold starts but increase idle carbon footprint. Depending on the function, idle carbon may even surpass (significantly) the carbon for execution (right plot). 
    }
   \label{fig:carbon}
\end{figure}

\textbf{Challenge II: Trade-off Between Responsiveness and Carbon Footprint.}
Keep-alive strategies create a fundamental conflict between responsiveness and sustainability. Using traces from the Huawei dataset~\cite{joosen2025serverless}, Fig. \ref{fig:carbon} demonstrates that increasing keep-alive timeouts sharply reduces cold starts, but monotonically increases the idle carbon footprint ($C_{idle}$).
A fixed timeout strategy cannot effectively balance this trade-off. 
This highlights the need for \textit{adaptive management} to dynamically adjust retention decisions based on specific function behavior. 

\begin{figure}[t]
    \centering
    \subfloat[Hourly carbon intensity profiles showing temporal variability.\protect\footnotemark
    ]{\label{subfig:cichange}\includegraphics[width=0.48\columnwidth]{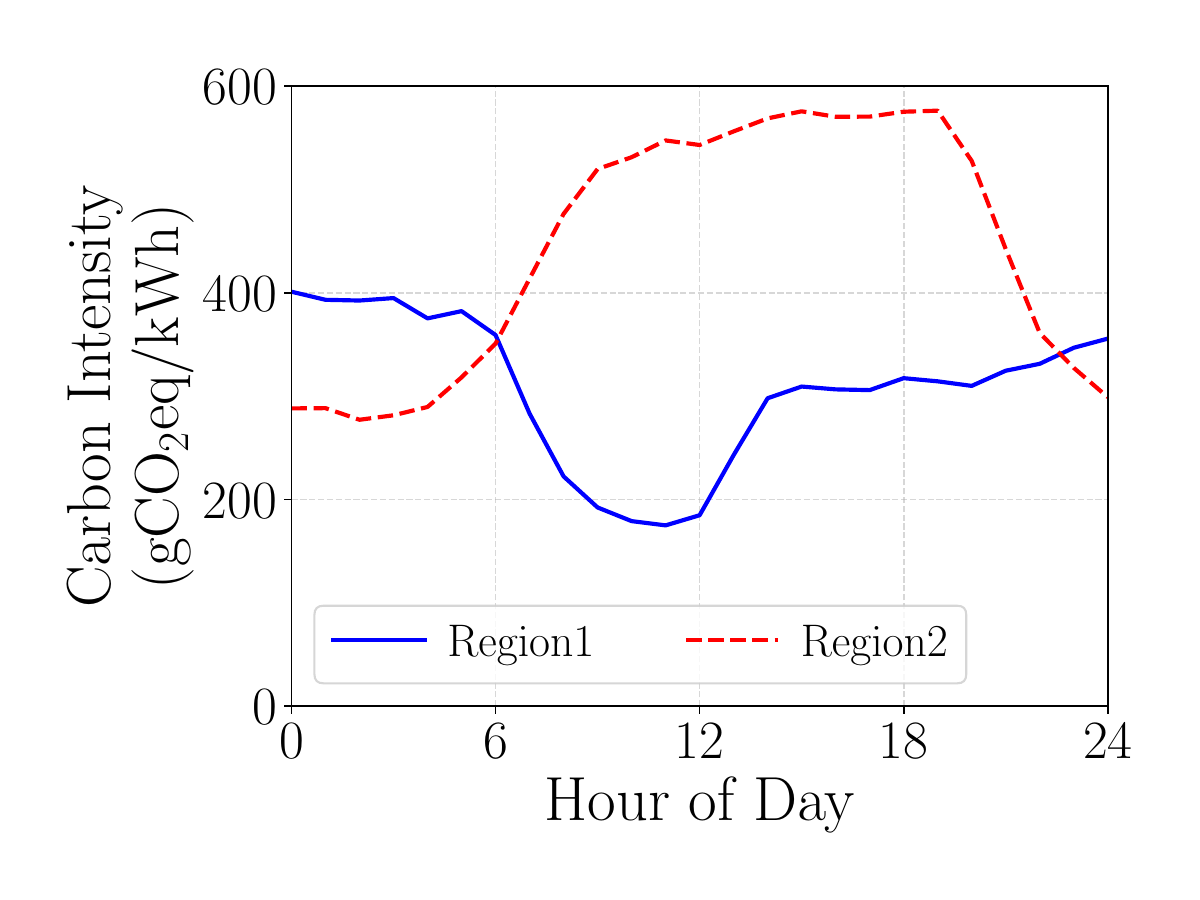}}
    \hfill
    \hspace{0.02\columnwidth}
    \subfloat[Function memory footprint CDF. The majority use less than 200MB.]{\label{subfig:memory}\includegraphics[width=0.48\columnwidth]{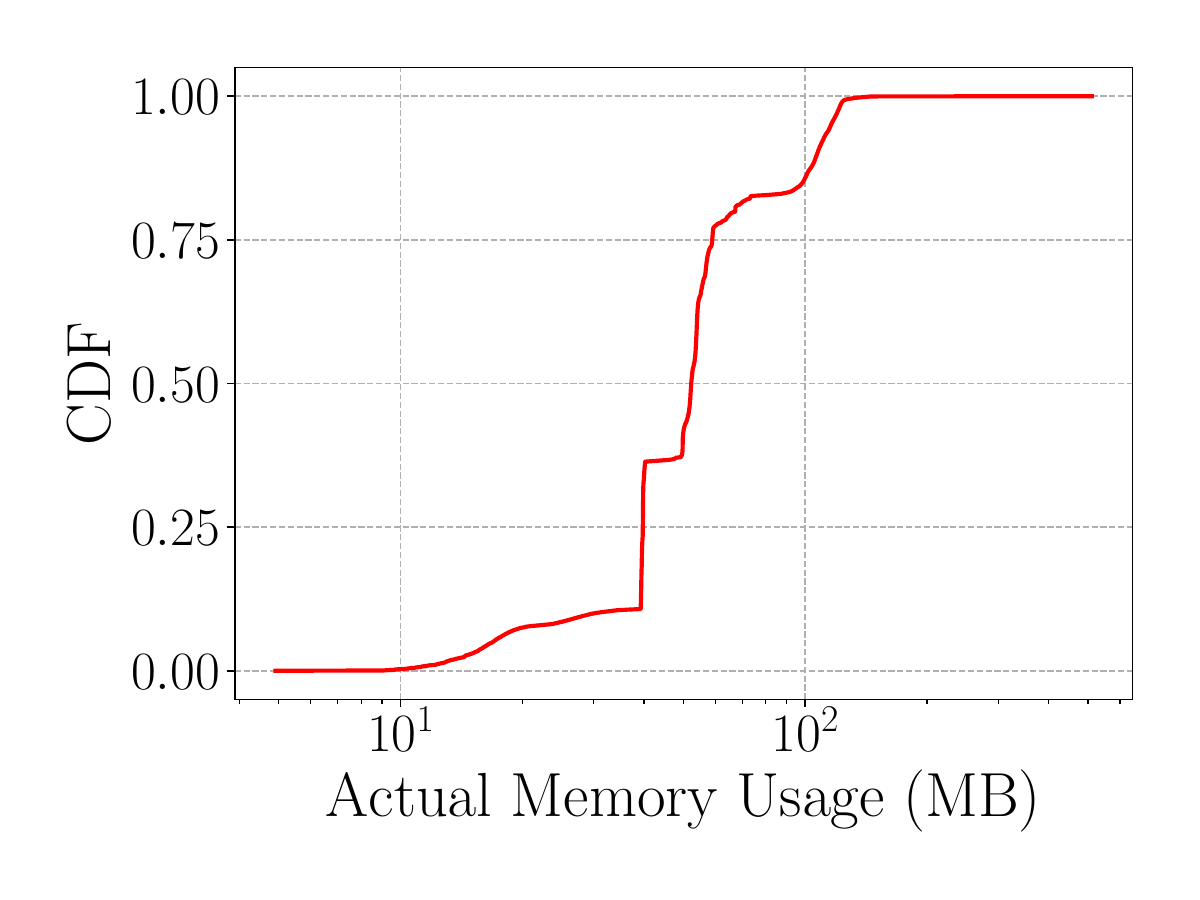}}
    \caption{Observed variability in carbon intensity of the electrical grid~\cite{electricitymaps}(a) and Huawei's function memory footprint (b).
    }
   \label{fig:ci-mem}
\end{figure}

\textbf{Challenge III: Temporal Variation in Carbon Intensity.}
Carbon footprint is heavily influenced by execution timing. Fig. \ref{subfig:cichange} shows hourly carbon intensity profiles from Electricity Maps~\cite{electricitymaps}, revealing substantial temporal variation (e.g., solar energy drops).
Static strategies miss opportunities to optimize for these fluctuations. \framename incorporates real-time carbon intensity to minimize idle costs during high-intensity hours while exploiting low-carbon periods to reduce cold starts.

\footnotetext{Region names are anonymized to preserve the double-blind review process.}
\textbf{Memory and Modeling Assumptions.}
Fig. \ref{subfig:memory} shows the CDF of the memory footprint of all function invocations in the Huawei dataset. The CDF indicates that more than 80\% of functions use less than 100MB, falling well below common allocation thresholds~\cite{aws_lambda,awsEC2m5,azure_functions}. Thus, we exclude memory constraints from the model formulation.
We further assume that function execution time (pod $execution$ phase) is independent of keep-alive decisions 
and that carbon intensity is provided as an external, real-time input~\cite{electricitymaps}.

\section{System Design}
\label{sec:system}
The \framename framework  optimizes pod management in serverless platforms by jointly considering the probability of cold start, cold start latency costs, and the carbon emissions of the idle keep-alive period. At its core is a reinforcement learning (RL) agent that learns to adapt keep-alive durations for each function execution. 

\subsection{\framename Overview}
\label{subsec:system-overiew}

\begin{figure}[htb]
    \centering
   \includegraphics[width=\columnwidth]{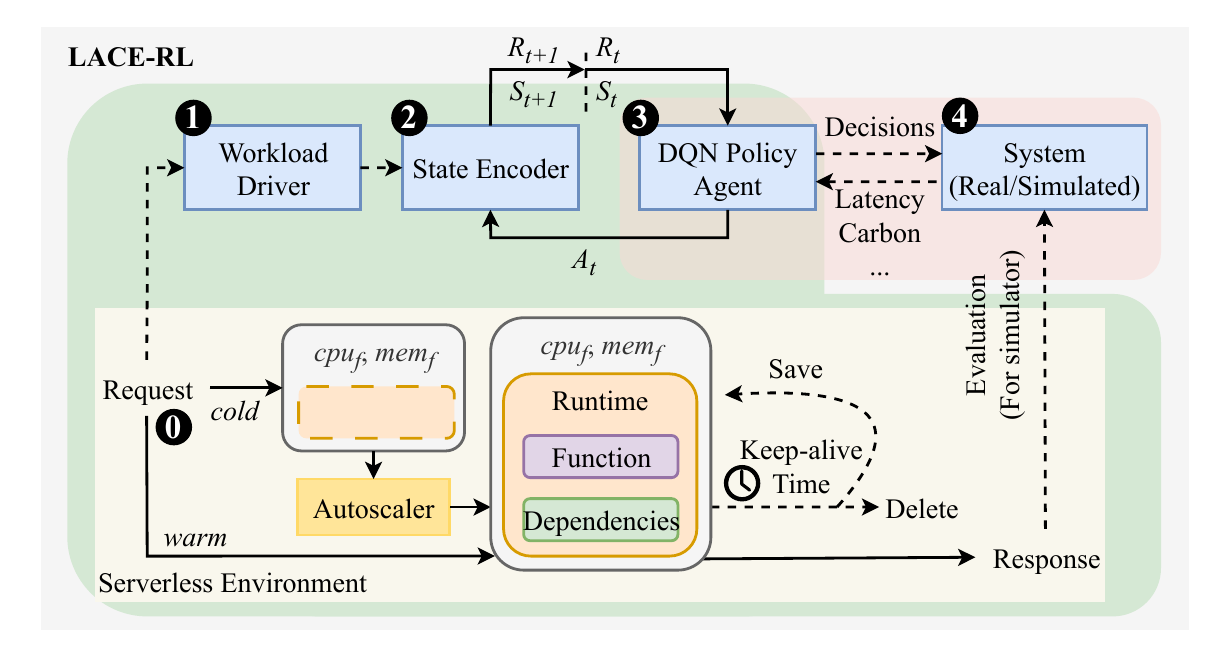}
    \caption{\framename overview.
    }
   \label{fig:overview}
\end{figure}

Fig. \ref{fig:overview} shows the four main components of \framename that collectively enable adaptive management: A \textit{Workload Driver}~\circled{1} that sequentially parses historical invocation traces and pod reuse states into the learning environment, a \textit{State Encoder}~\circled{2} that transforms management context into (normalized) feature vectors, a \textit{DQN Policy Agent}~\circled{3} that selects actions based on observed state and user-defined tradeoffs, and a \textit{Real System or Simulator}~\circled{4} that applies decisions and reports performance outcomes. 
Large-scale testing of middleware / system software on production-grade serverless platforms is infeasible due to limited control. Our current evaluation is conducted within the trace-driven simulator (as commonly adopted in state-of-the-art work~\cite{jiang2024ecolife, lee2024spes}) for reproducibility and scalability. The architecture is fully compatible with online deployment, where the agent can make real-time decisions based on live system states.

\textbf{Workload Driver.} The workload driver serves as the backbone of \framenames offline training loop. It parses the real-world trace of function invocations -- each with timestamps, resource requirements, pod identifiers, and function metadata -- and streams them sequentially into the environment. This setup enables the policy agent to observe realistic invocation bursts and temporal reuse patterns, which are crucial for modeling cold-start dynamics.

\textbf{State Encoder.} For each function invocation $i$ at time $t$, the state encoder constructs a $d$-dim feature vector $S_t \in \mathbb{R}^d$ representing the current context. Specifically, $S_t$ includes: the reuse probability $p_{k}$ of its pod estimated using a historical window $W$ over different keep-alive durations $k \in \mathcal{K}_{keep}$, memory and CPU requests $res_{k}^{mem}$, $res_{k}^{cpu}$, cold start latency $L_{i}^{cold}$ for invocation $i$, real-time carbon intensity $CI_{t}$ at timestamp $t$, and user-defined tradeoff weights $\lambda_{carbon} \in [0,1]$ (as commonly done for tunable parameters \cite{jiang2024ecolife,li2023clover}). 
We log-normalize long-tailed latency features and standardize energy features using training-set statistics. The state vector $S_t$ is the DQN input, allowing the agent to jointly reason about workload behavior, sustainability constraints, and user preferences.

\textbf{DQN Policy Agent.} The agent uses a Deep Q-Network (DQN), a neural network–based approximation of the Q-learning algorithm in RL to map state vectors to actions. Each action is a discrete keep-alive timeout $t$. The agent is trained through reward feedback and updates its policy through temporal-difference learning, enabling it to dynamically balance cold starts, carbon cost, and latency. The formulation of the RL model in \framename is introduced in Section~\ref{subsec:rl-formulation}.

The system forms a closed-loop learning environment. During offline training of the RL agent, the workload driver provides data, and the encoder prepares state features for the policy agent to learn how to select actions. After the agent is trained, it is applied in a per-invocation-triggered way, at just a microsecond cost, to do online inference. The decisions can either be applied to the real system or evaluated using a simulator.

\subsection{Carbon-Latency Tradeoff via Keep-Alive Tuning}
Serverless computing environments commonly face a trade-off between reducing cold start latency and minimizing carbon emissions. \framename models this tradeoff explicitly by allowing the RL agent to select discrete keep-alive decisions for each pod. 

For each function invocation $i$ at time $t$, the agent chooses a keep-alive timeout $k$ from a predefined set $\mathcal{K}_{keep}$. A higher value of $k$ reduces the cold start probability $p_{k}$, but increases idle carbon emissions due to prolonged pod retention. To capture this trade-off, we define two cost terms:
\begin{enumerate}
    \item $C_{\text{cold}}(k) = (1-p_{k}) \cdot L_{i}^{cold}$, where $L_{i}^{cold}$ is the cold start latency for invocation $i$. This term quantifies the \textit{expected latency penalty} a user experiences due to cold starts, weighted by the likelihood that the pod expires before reuse.
    \item $C_{\text{carbon}}(k) = E_{i}^{\text{idle}} \cdot CI_t$, where $E_{i}^{\text{idle}}$ is the idle energy consumption associated with keeping the pod alive for $k$ seconds (as introduced in Section \ref{subsec:energymodel}) and $CI_t$ is the carbon intensity at time $t$. This cost term reflects the \textit{environmental impact} of keeping pods warm during idle periods.
\end{enumerate}
Based on the two cost terms, the reward assigned to each action is defined as follows:
\begin{equation}
\label{eq:reward-core}
    R= -[ (1-\lambda_{carbon}) \cdot C_{\text{cold}}(k) + \lambda_{carbon} \cdot C_{\text{carbon}}(k)]
\end{equation}
The reward formulation in Eq. (\ref{eq:reward-core}) allows \framename to dynamically trade between performance and sustainability based on the tunable preference weight $\lambda_{carbon} \in [0,1]$. When $\lambda_{carbon}$ is close to 0, the agent prioritizes minimizing latency. On the other hand, when $\lambda_{carbon}$ is close to 1, the agent of \framename emphasizes carbon efficiency. Intermediate values produce hybrid strategies that interpolate between the two objectives.

\subsection{Reinforcement-Learning Model Formulation}
\label{subsec:rl-formulation}
\framename formulates function management as a Markov Decision Process (MDP), where the agent learns to optimize deployment based on the encoded system context.
At each invocation, the action sets the pod expiration time, which (i) determines for how long idle energy may be consumed until reuse or pod expiration and (ii) affects whether subsequent invocations will find the pod warm or will suffer a cold-start.

\textbf{Why RL and Why DQN.} We employ reinforcement learning in \framename because keep-alive tuning is a sequential decision problem characterized by delayed rewards. Unlike one-shot optimization, a decision made now affects resource availability and cold start probability in the future. Analytical solutions and static heuristics fail to capture the complex, non-linear interplay between reuse dynamics and real-time carbon intensity. Furthermore, Supervised Learning is unsuitable as there is no ground truth for the ``correct" timeout; the optimal policy must be learned through exploration and interaction.

We adopt the Deep Q-Network (DQN) algorithm to approximate the action-value function over a continuous state space. We select DQN over Tabular Q-learning, which cannot scale to high-dimensional continuous states, and over on-policy methods (e.g., PPO), which are less sample-efficient for trace-driven training. 
DQN is an off-policy algorithm, allowing \framename to learn effectively from the replay of historical experiences. Finally, our specific DQN implementation utilizes a lightweight architecture yielding microsecond-level inference overhead suitable for real-time deployment (Sec. \ref{subsec:inference}).

\textbf{MDP Formulation.} As shown in Fig. \ref{fig:overview}, at each decision point $t$, the agent observes a state vector $S_t \in \mathbb{R}^d$ that encodes reuse probabilities across keep-alive durations based on historical data, normalized memory and CPU usage, estimated cold start latency, real-time carbon intensity and trade-off weight $\lambda_{carbon}$. The agent selects an action $a_t \in \mathcal{A}$ to optimize cumulative long-term rewards. The environment provides feedback based on the real-world trace through a scalar reward $R_t$ and transitions to the next state $S_{t+1}$. This sequential interaction fits within the standard MDP framework defined by a tuple $(\mathcal{S}, \mathcal{A}, \mathcal{P}, \mathcal{R},\gamma)$. Among them, $\mathcal{P}$ are the transition dynamics, implicitly encoded via the trace-driven simulator, and $\gamma \in (0,1]$ is the discount factor for the accumulation of rewards in each episode. 
$\mathcal{S}$, $\mathcal{A}$, and $\mathcal{R}$ correspond to the state space, action space, and reward function and are defined as:

\textbf{State Space.} Using the notation given in Section~\ref{subsec:system-overiew}, $S_t$ is defined as:
\begin{equation}
    S_t = [p_{k_1}, \ldots, p_{k_n}, r_i^{\text{mem}}, r_i^{\text{cpu}}, L_i^{\text{cold}}, CI_t, \lambda_{\text{carbon}}].
\end{equation}
This compact state representation ensures that workload dynamics and preferred weights are available to the agent at each decision point.
At inference time, the agent selects actions per invocation using only the observed state $S_t$ without access to future invocations.

\textbf{Action Space.} The action space $\mathcal{A}$ is the set of keep-alive candidates $\mathcal{K}_{keep}$. Each action is a scalar $a_t$. In practice, all actions are mapped to discrete indices for efficient computation and lookup of $Q$ values.

\textbf{Reward Function.} Rewards are defined in Eq. (\ref{eq:reward-core}) to integrate cold start latency, carbon emissions from idle energy, and pod reuse probability, evaluated using real trace data, temporal carbon intensity, and network latency.

\textbf{DQN and Strategy Learning.} \framename employs Deep Q-Learning (DQN) to learn optimal management strategies. A fully connected neural network is used to  approximate the action-value function $Q(s,a;\theta)$, where $\theta$ represents the collection of learnable parameters. 
The input of DQN is the full state vector $S_t$ and the output is the Q-values for all discrete actions $|\mathcal{A}|$. Experience replay and target networks are used to stabilize training. 
The DQN training steps are:
\begin{itemize}[leftmargin=10pt]
    \item \textbf{Experience Replay.} Transitions ($s_t$, $a_t$, $R_t$, $s_{t+1}$) are stored in a buffer and randomly sampled for training.
    \item \textbf{Target Network.} A separate target network $Q^{\prime}$ is used to compute stable targets during training and is updated periodically by copying weights from the primary Q-network.
    \item \textbf{Epsilon-greedy Policy.} The agent selects random actions with probability $\epsilon$ to explore the environment. $\epsilon$ decays over time to promote convergence.
    \item \textbf{Loss.} The network minimizes the squared Temporal Difference (TD) loss $\mathcal{L}$ during the training, defined as: \begin{equation}
        \mathcal{L}(\theta)=\mathbb{E}_{(s, a, r, s^{\prime})}[(r+\gamma \max _{a^{\prime}} Q^{\prime}\left(s^{\prime}, a^{\prime}\right)-Q(s, a ; \theta))^2]
    \end{equation}
\end{itemize}

\textbf{User-tunable Preference.} As discussed in Section~\ref{subsec:system-overiew}, \framename explicitly embeds user-tunable parameter $\lambda_{carbon}$ in the state vector. This allows the Q-network to learn a preference-conditioned strategy that adapts dynamically during inference. As a result, users can specify tradeoffs at run-time without retraining. This improves the flexibility of deployment in diverse or evolving operational settings.

\section{Evaluation}
\label{sec:evaluation}

Our goal is to assess how effectively \framename co-optimizes cold starts and carbon emissions while maintaining low end-to-end function invocation latency. We evaluate \framename from five perspectives: (1) performance on general workloads (Section~\ref{subsec:genraltest}), (2) robustness under challenging, high-cold-start-latency scenarios (Section~\ref{subsec:taildistr}), (3) comparison against an Oracle-based policy (Section~\ref{subsec:oracle}), (4) inference cost and training overhead (Section~\ref{subsec:inference}), and (5) sensitivity analysis of tunable parameters (Section~\ref{subsec:sensitivity}).

\subsection{Experimental Methodology}
\label{subsec:exp-setup}

We utilize a trace-driven simulation approach. To ensure this simulator is realistic, we first conduct real-system characterization to validate our simulation parameter assumptions.

\subsubsection{Model and Parameters Validation via Real-System Measurements}
\label{subsubsec:profiling}
To validate the model and parameters adopted in \framename, we profile a deployment using FunctionBench~\cite{kim2019functionbench}, a widely accepted FaaS workload benchmark. Since the Huawei trace provides only hashed function IDs without source code, direct profiling of trace workloads is infeasible. We therefore utilize this controlled benchmark to validate the energy accounting factors used by our simulator.

We characterize energy consumption across the three primary serverless phases: cold start, active execution, and keep-alive (warm) retention. Experiments are conducted on an HPE ProLiant DL385 (Gen10 Plus v2) server equipped with dual AMD EPYC 7513 processors (32 cores per socket, 64 cores total) and 256 GB of memory. We deploy Knative~\cite{knative2025} on Kubernetes~\cite{brewer2015kubernetes} and utilize Kepler~\cite{kepler2025} for energy monitoring. Kepler reports node-level active and idle energy at the CPU package level.

To ensure accurate per-function accounting, each function is profiled in isolation. Most pods are configured to request a single CPU core ($c_i=1$). However, to reflect realistic deployment patterns for compute-intensive workloads, functions such as Matrix Multiplication and Linpack (implemented using NumPy) are allowed to execute using multiple cores ($c_i > 1$).

For every run, we calculate the total per-pod energy by summing two components: the direct active energy and the allocated idle baseline. This attribution logic applies consistently across all phases (cold start, compute, and keep-alive):
\begin{enumerate}[label=\roman*), leftmargin=10pt]
    \item \textit{Active Component:} We integrate the node-level active power reported by Kepler over the duration of the phase. Since only one pod is active during profiling, this energy is attributed entirely to the target function.
    \item \textit{Idle Baseline:} The baseline static energy of the node is attributed proportionally to the resources reserved by the pod. For a pod reserving $c_i$ cores on a node with capacity $C=64$, the allocated idle energy is calculated as $E^i_{idle} = (c_i / C) \cdot E_{node\_idle}$. We should note that the average idle power consumption remains stable across different phases of the execution life of each pod. 
    This proportional accounting ensures that heavier workloads (with $c_i > 1$) bear a larger share of the static power costs, while lightweight functions are not penalized for the server's total idle power. We focus on CPU-based attribution because memory usage across all benchmarks ranges from only 42 MB to 275 MB (Table \ref{tab:energy-profiling}). This represents at most 0.11\% of system memory, confirming that consolidation is CPU-bound rather than memory-bound.
\end{enumerate}

Table \ref{tab:energy-profiling} summarizes the results. We adopt the original benchmark names from FunctionBench~\cite{kim2019functionbench}. 
We observe significant cross-function diversity. Multicore compute-intensive kernels (e.g., Linpack) exhibit substantially higher total power due to both high active utilization and the scaled idle attribution described above. 

For the majority of workloads, cold-start energy is small relative to the compute and keep-alive energy due to the short duration of the cold-start phase. The main outliers are Image/Video Processing and Image Classification, whose cold-start times (and corresponding active energies) are markedly longer due to heavier initialization overhead (e.g., library dependencies and model weight loading). On average, the cold-start phase duration is a good predictor for the respective energy cost.

The keep-alive-to-compute power ratio ($\lambda_{idle}$) spans 0.21--0.83, indicating that warm retention consumes a significant fraction of active power. In our simulation, we set $\lambda_{idle}=0.2$. This is a \textbf{conservative choice} relative to the measured values. Applying a larger $\lambda_{idle}$ would further increase the modeled cost of idle pod retention, thereby strengthening the motivation for the adaptive strategy proposed by \framename. A sensitivity analysis of $\lambda_{idle}$ is illustrated in Sec. \ref{subsec:sensitivity}.

\begin{table*}[t]
\centering
\resizebox{\textwidth}{!}{
\begin{tabular}{lcccccccccc}
\toprule
\makecell[l]{Benchmark\\Function Name} &
\makecell{Input\\Size / Dimension} &
\makecell{Memory\\Consumption} &
\makecell{Cold Start\\Time (ms)} &
\makecell{Compute\\Time (ms)} &
\makecell{Cold-start\\Active\\Energy (J)} &
\makecell{Compute-phase\\Active\\Energy (J)} &
\makecell{1-min\\Keep-alive-phase\\Active Energy (J)} &
\makecell{Per-pod\\Compute-Phase\\Total Power (W)} &
\makecell{Per-pod\\Keep-alive-phase\\Total Power (W)} &
\makecell{Lambda Idle\\Total Power Ratio\\(Keep-alive/Compute)} \\
\midrule
Float Operations    & 10,000,000 & 44 MB    & 112.2    & 3340.86 & 0.94 & 15.08 &	78.29 & 6.37   & 3.19 & \textbf{0.50} \\
MatMul              & 10,000 & 95 MB                & 166.5    & 2393.41 & 0.27	&144.41	& 76.98 & 86.64 & 28.89 & \textbf{0.33} \\
Linpack             & 100,000 & 97 MB               & 76.33    & 6401.45 & 0.7	&436.9	& 92.4 & 147.29 & 70.82 & \textbf{0.48} \\
Image Processing  & 28.4 MB & 68 MB               & 2441.68 & 6761.82 & 11.13	& 20.69	& 81.6 & 4.98   & 3.21 & \textbf{0.64} \\
Video Processing  & 742 KB & 233 MB               & 12414.77 & 2403.04 & 19.05	& 6.82	& 72.68 & 4.65   & 3.03 & \textbf{0.65} \\
Chameleon           & [500,100]   & 57 MB            & 71.6     & 249.52   & 0.52	& 1.84	& 81.1 & 9.27   & 3.14 & \textbf{0.34} \\
pyaes               & 200 iterations   & 42 MB       & 563.17   & 1567.58 & 3.41	& 6.34	& 66.78 & 6.02   & 2.87 & \textbf{0.48} \\
Feature Extractor & 30.5 MB   & 133 MB             & 109.31   & 2323.78 & 0.15	& 10.40	& 75.04 & 6.33   & 3.06 & \textbf{0.48} \\
Model Training         & 15.23 MB & 172 MB           & 115.58   & 2485.6   & 2.96	& 31.66	& 79.2 & 14.56 & 3.12 & \textbf{0.21} \\
Classification Image & 28.4 MB & 275 MB            & 8642.95 & 1591.42 & 21.39	& 2.96	& 71.42 & 3.68   & 3.05 & \textbf{0.83} \\
\bottomrule
\end{tabular}
}
\caption{Energy profiling of serverless function pods under cold start, compute, and keep-alive phases.}
\label{tab:energy-profiling}
\end{table*}

\subsubsection{Dataset and Trace Processing}
We evaluate \framename using the Huawei Cloud Trace~\cite{joosen2025serverless}, which contains fine-grained logs of invocation timestamps, pod assignments, runtime types, and cold start latencies. The trace records over 300 million requests and cold start logs for more than 1,500 unique functions, providing a rich dataset for model training and performance validation. To construct a realistic evaluation environment, we apply the following pre-processing steps:

\begin{itemize}[leftmargin=10pt]
    \item \textbf{Training-Validation-Testing Data Partition.} Records are grouped by pod identifier (\texttt{podID}) to preserve temporal reuse patterns. The resulting dataset is split into 80\% training, 10\% validation, and 10\% testing.
    \item \textbf{Cold Start Profiling.} We utilize the cold start logs to generate latency estimates based on trigger type and runtime (e.g., Python, Custom). A lookup table, derived from the training data, provides the expected cold start latency $L_i^{cold}$ for each invocation, allowing the simulator to emulate system delays during cold starts.
    \item \textbf{Resource Usage Integration.} We join the request-level logs with cold start traces and metadata to produce a complete invocation stream containing CPU/memory usage, timestamps, and cold start latency profiles. These features serve as the state input for the RL agent.
\end{itemize}

\subsubsection{Energy and Carbon Modeling}
Incorporating the structural findings from our characterization, we implement a trace-driven simulator that models a homogeneous cluster of serverless nodes. 
To ensure our absolute carbon estimates reflect typical public cloud deployments, we model a cluster of standard
m5-series EC2 instances (32 logical cores per 128 GB DRAM)~\cite{awsEC2m5}. Energy estimates per active core and per MB of memory are derived from the Thermal Design Power (TDP) and benchmarks of the Intel Xeon Platinum 8275CL processor~\cite{intelXeon}, representing the class of CPUs powering the m5 instance family. 
To model the CO\textsubscript{2} impact, we incorporate real-world carbon intensity traces from Electricity Maps~\cite{electricitymaps} with an hourly sampling rate. For each invocation, the total carbon cost is computed by weighting the execution and keep-alive energy by the carbon intensity at the specific time of day. This setup allows \framename to adapt keep-alive choices to changing carbon conditions.

\subsubsection{RL Model Configuration}
The action space of RL agent consists of discrete keep-alive durations: $\{1, 5, 10, 30, 60\}$ seconds. These values capture the empirical distribution of average reuse intervals (corresponding roughly to the 10th, 50th, 75th, and 90th percentiles) and include the standard 60s timeout used by Huawei~\cite{joosen2025serverless} to ensure compatibility with real-world practice.

\textbf{Training and Inference.} The agent is trained offline using a replay buffer of 10,000 samples, an input batch size of 64, a learning rate of 0.001, and a reward discount factor $\gamma=0.99$. 
Epsilon-greedy exploration starts at $\epsilon=1.0$ and decays at a rate of 0.95 per episode to a minimum of 0.05. 
Stable convergence is observed after approximately 300 episodes.

Once trained, the agent is evaluated on a \textit{testing set} unseen during training. For each request, the simulator feeds the encoded current state to the DQN to select the optimal keep-alive time. In line with our earlier analysis (Fig. \ref{subfig:cslatencycdf}), where cold start latency spans from sub-second to more than 10s with a long-tailed distribution, we structure evaluation around two workload sets:
\begin{itemize}[leftmargin=10pt]
    \item \textbf{General:} The entire testing set (all invocations).
    \item \textbf{Long-tailed:} A subset containing only the high-latency functions of General (long tail of the latency distribution), which are heavily affected by cold starts.
\end{itemize}

\subsubsection{Baseline Strategies}
We compare \framename against four strategies: 
\begin{itemize}[leftmargin=10pt]
    \item \textbf{Latency-Minimizing (\textit{Latency-Min}):} Minimizes expected cold starts ($C_{cold}$) regardless of energy cost.
    \item \textbf{Carbon-Minimizing (\textit{Carbon-Min}):} Minimizes keep-alive duration to strictly reduce idle carbon ($C_{carbon}$), often at the cost of high latency.
    \item \textbf{Huawei:} A static strategy using a fixed 60s keep-alive timeout, consistent with the state-of-the-practice platform configuration~\cite{joosen2025serverless}.
    \item \textbf{DPSO (EcoLife~\cite{jiang2024ecolife}):} A state-of-the-art carbon-aware metaheuristic using Particle Swarm Optimization to jointly optimize keep-alive durations and function placement.
\end{itemize}

\subsubsection{Evaluation Metrics}
We evaluate \framename using standard system metrics: i) \textbf{Cold Start Count}, ii) \textbf{Average End-to-end Latency} including cold start, execution, and a fixed network latency profiled via AWS CloudPing~\cite{cloudping}\footnote{
     Since the evaluation is restricted to a single-site setting, network latency is treated as a constant offset and does not affect adaptive keep-alive decisions.}, iii) \textbf{Keep-alive Carbon} resulting from idle pods kept alive for invocations, and iv) \textbf{Total Carbon Emissions} including both execution and keep-alive carbon emissions.
    
To enable unified trade-off analysis, we introduce two composite metrics inspired by the Energy-Delay Product~\cite{laros2012energy, valentini2013overview} used to evaluate energy-performance trade-offs in the HPC domain (for both metrics, lower values indicate better efficiency):
\begin{itemize}[leftmargin=10pt]
    \item \textbf{Latency–Carbon Product (LCP):} The product of average end-to-end latency and total carbon emissions. It quantifies the overall efficiency of the system, trading off responsiveness against environmental impact.
    \item \textbf{Idle Reuse Inefficiency (IRI):} The product of cold start count and keep-alive carbon. This quantifies the wastefulness of idle resources that fail to prevent cold starts.
\end{itemize}

\subsection{Performance on General Testing Set}
\label{subsec:genraltest}

\begin{figure}[t]
    \centering
    \subfloat[Cold Start Counts.]{\label{subfig:latcscount}\includegraphics[width=0.5\columnwidth]{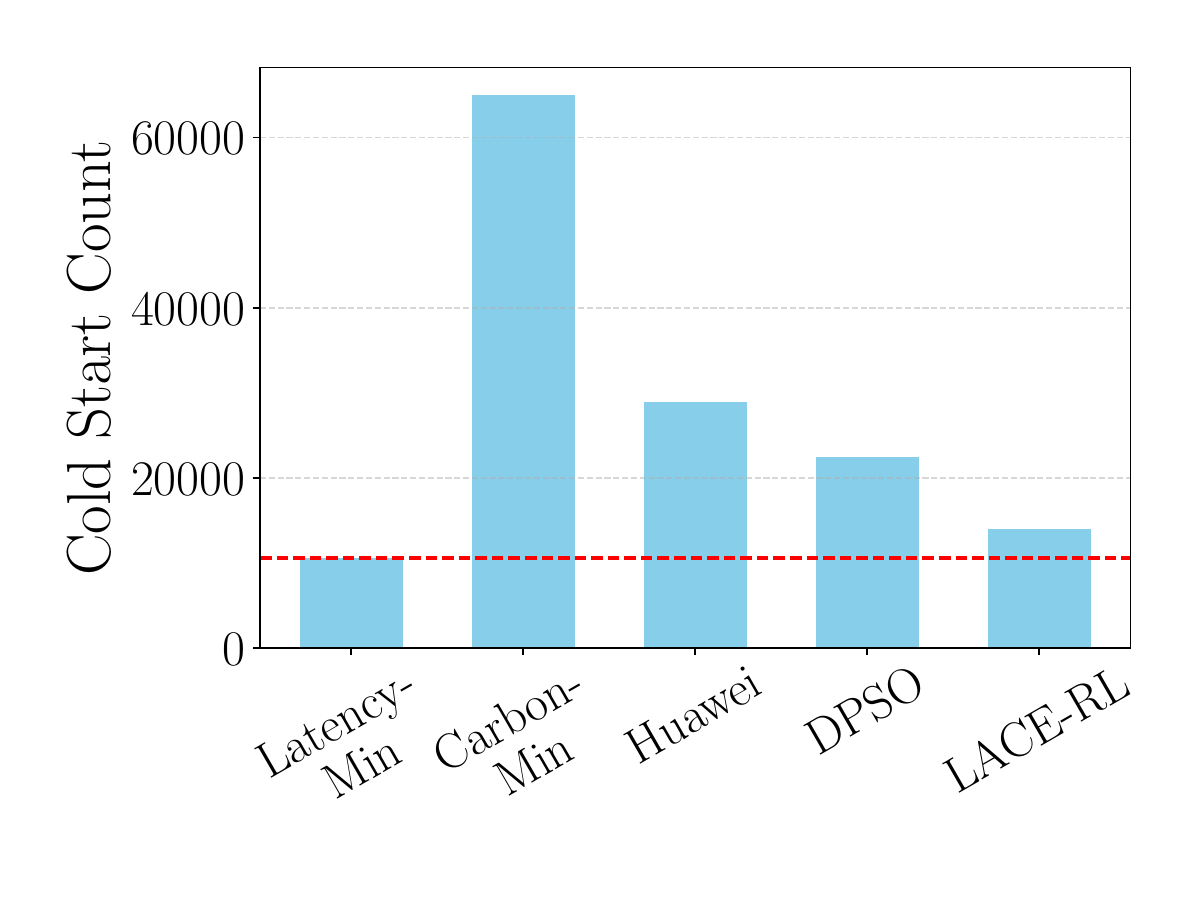}}
    \hfill
    \subfloat[End-to-end latency.]{\label{subfig:e2elatency}\includegraphics[width=0.5\columnwidth]{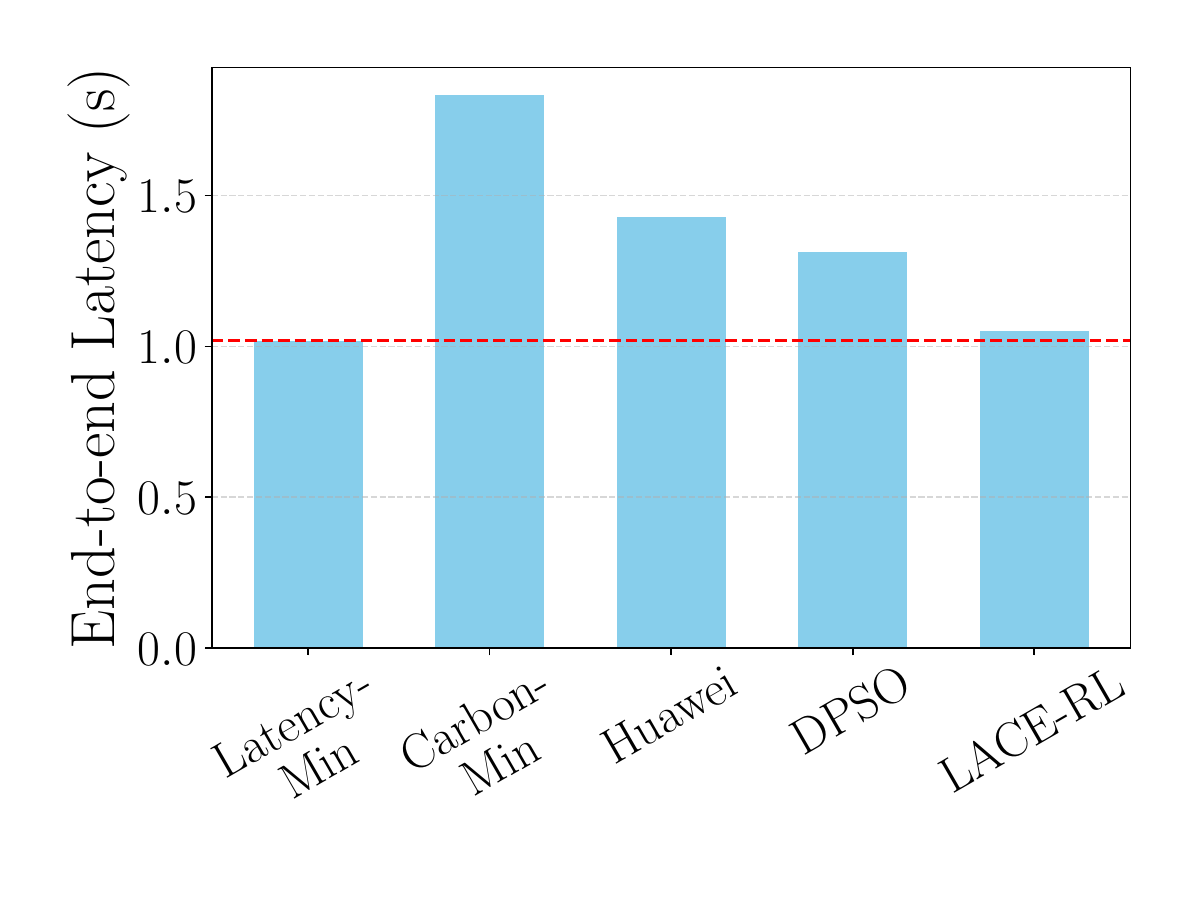}}
    \vfill
    \centering
    \subfloat[Keep-alive Carbon.]{\label{subfig:latkacarbon}\includegraphics[width=0.5\columnwidth]{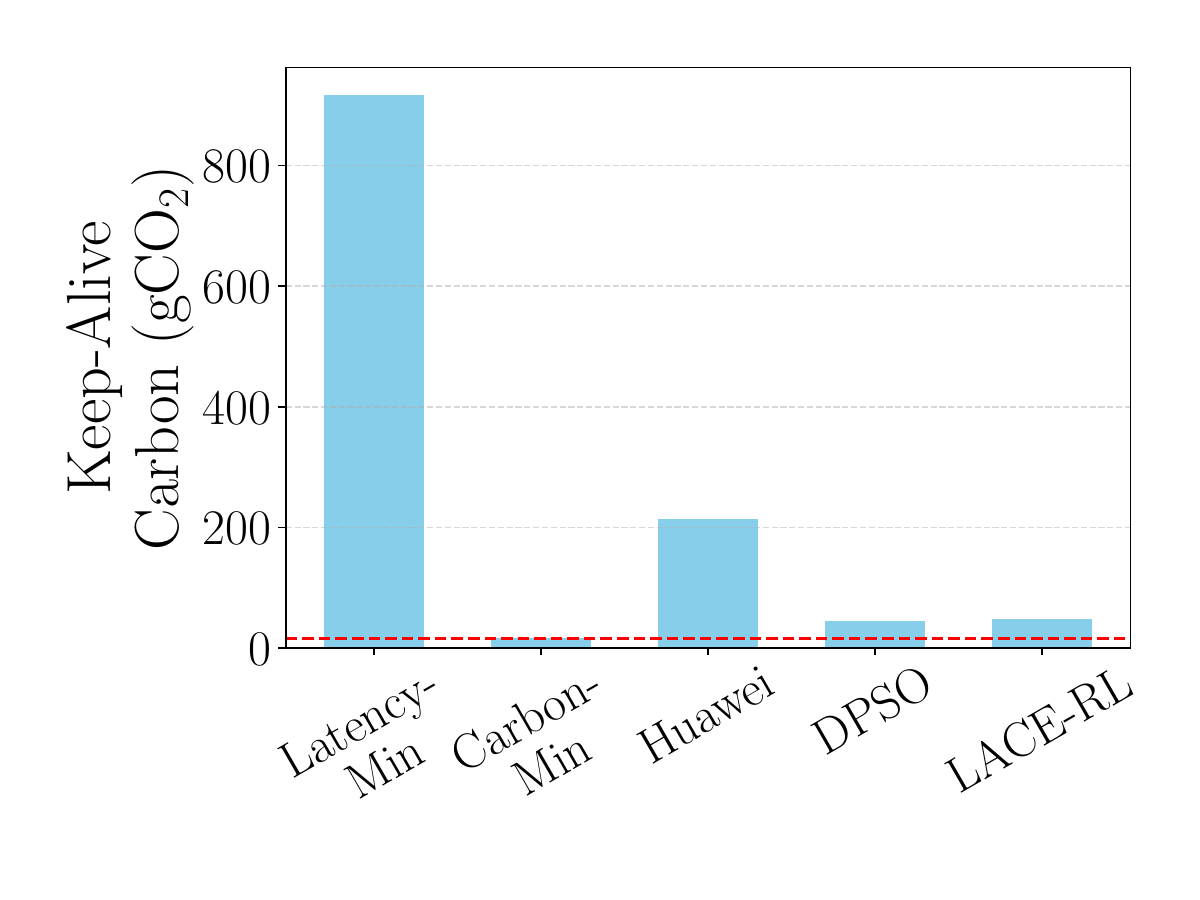}}
    \hfill
    \subfloat[Total Carbon Footprint.]{\label{subfig:lattotcarbon}\includegraphics[width=0.5\columnwidth]{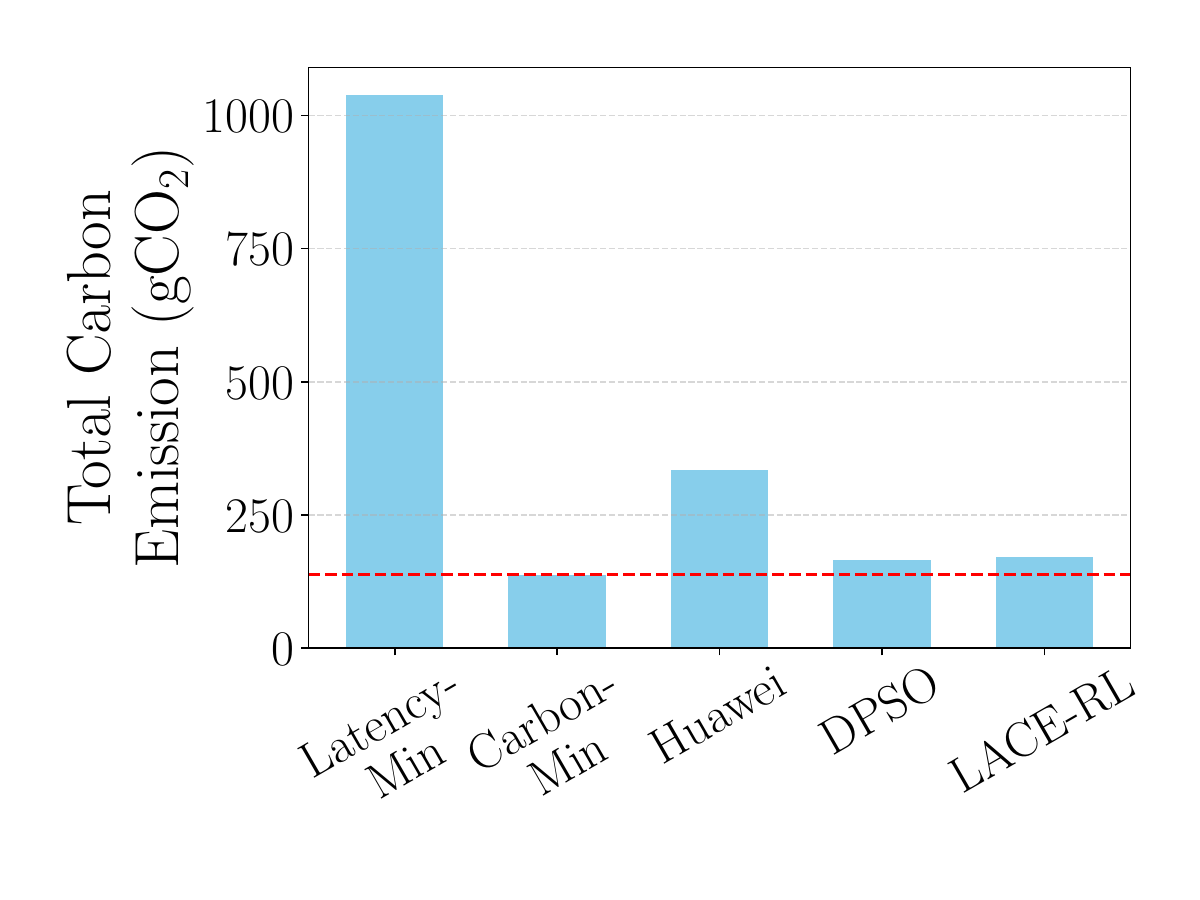}}
    \caption{General testing workload results across key metrics. The dashed line shows the optimal value.}
    \label{fig:avglat}
\end{figure}

\begin{figure}[t]
    \centering
   \includegraphics[width=\columnwidth]{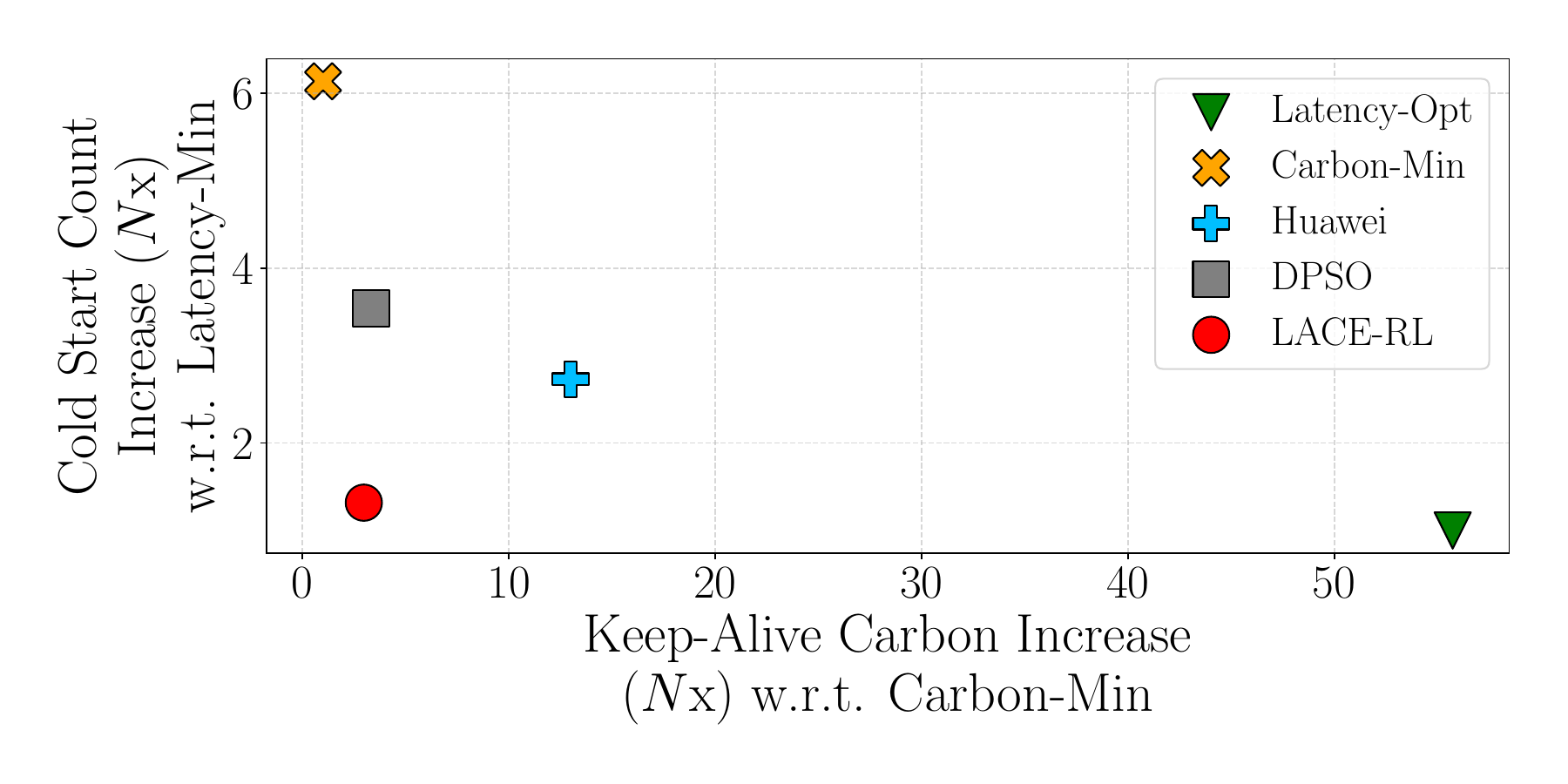}
    \caption{For the General workload \framename achieves the best tradeoff between cold starts and keep-alive carbon.}
   \label{fig:avglatrelative}
\end{figure}

We begin with the General workload comprising 99,140 invocations from the Huawei Cloud Trace~\cite{joosen2025serverless}. Fig.~\ref{fig:avglat} presents the absolute metric comparisons, while Fig.~\ref{fig:avglatrelative} highlights relative trade-offs. Together, they demonstrate that \framename achieves a delicate balance between responsiveness and sustainability.

\textbf{Cold Start Counts.} Fig. \ref{subfig:latcscount} shows that \framename reduces cold starts to 13,958. This represents a 51.7\% reduction compared to the \textit{Huawei} default (28,891) and a 37.9\% reduction compared to \textit{DPSO} (22,485). While \textit{Latency-Min} achieves a slightly lower count (10,588), it does so by indiscriminately prolonging keep-alive durations, leading to massive energy waste. \framename achieves its efficiency by selectively retaining pods based on predicted reuse likelihood, avoiding over-provisioning for low-impact functions.

\textbf{End-to-end Latency.} Fig.~\ref{subfig:e2elatency} shows that \framename achieves an average end-to-end latency of 1.05s. This effectively matches the performance of \textit{Latency-Min} (1.02s) while outperforming \textit{DPSO} (1.31s), \textit{Huawei} (1.43s), and \textit{Carbon-Min} (1.83s). By dynamically adapting to burst patterns and cold-start risks, \framename maintains high responsiveness without the excessive idle costs of greedy strategies.

\textbf{Carbon Metrics.} In terms of sustainability, Fig.~\ref{subfig:latkacarbon} shows that \framename emits only 48.93g CO\textsubscript{2} for keep-alive retention. This is drastically lower than \textit{Huawei} (213.52g) and \textit{Latency-Min} (916.67g), and comparable to \textit{DPSO} (44.73g). Total carbon emissions (Fig.~\ref{subfig:lattotcarbon}) follow the same trend, with \framename emitting 121.23g. Notably, \framename matches the carbon efficiency of \textit{DPSO} while delivering far superior latency and cold start performance, reflecting the benefits of finer-grained, per-invocation decision-making over population-based heuristics.  

\textbf{Trade-Off Evaluation.} Fig. \ref{fig:avglatrelative} visualizes the normalized trade-off, plotting cold start increase (vs. \textit{Latency-Min}) against keep-alive carbon increase (vs. \textit{Carbon-Min}). An ideal scheduler would lie in the bottom-left corner. \framename is positioned closest to this optimal point, confirming it explores the latency–carbon space more effectively than all baselines.

\begin{figure}[t]
    \centering
    \subfloat[Latency–Carbon Product (LCP).]{\label{subfig:lcp}\includegraphics[width=0.48\columnwidth]{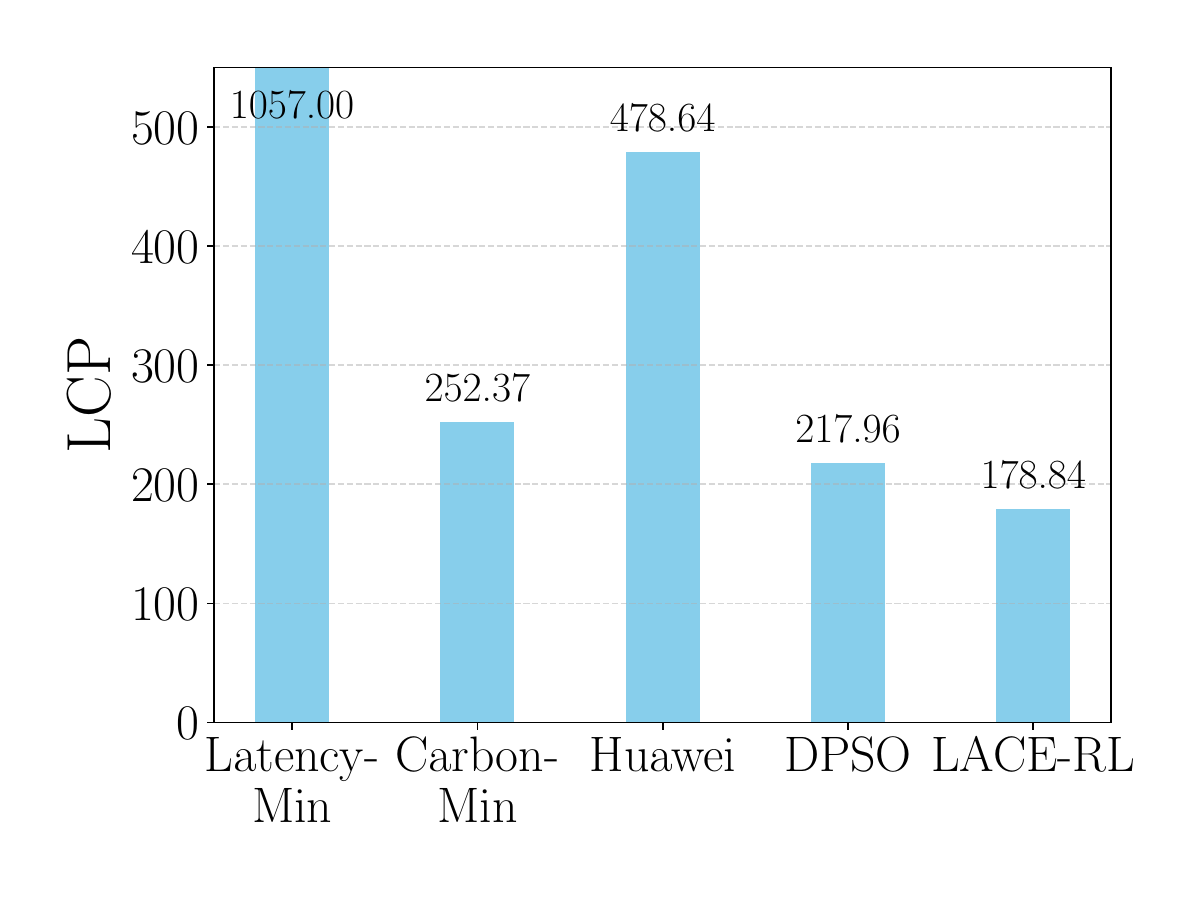}}
    \hspace{0.02\columnwidth}
    \subfloat[Idle Reuse Inefficiency (IRI).]{\label{subfig:iri}\includegraphics[width=0.48\columnwidth]{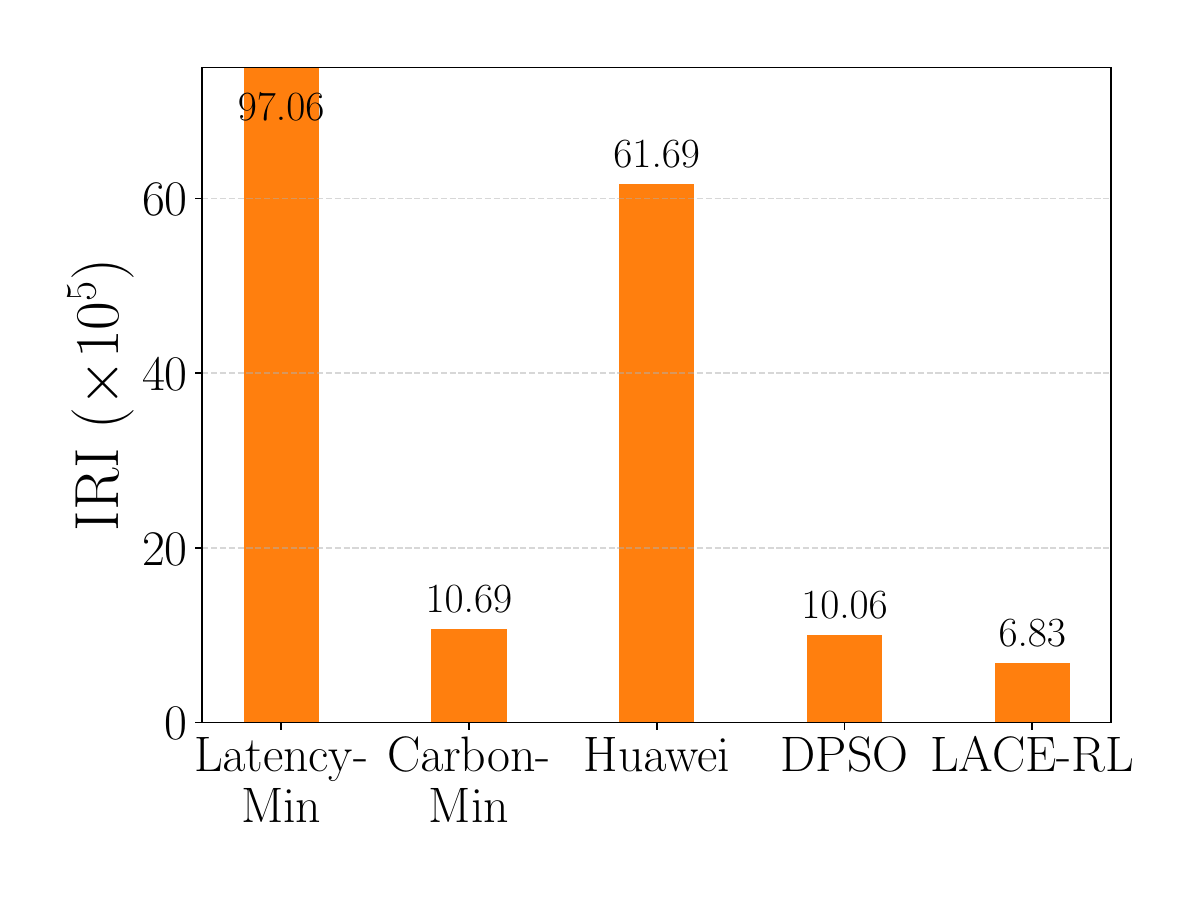}}
    \caption{Comparison based on the composite metrics (lower is better). \framename always outperforms the other strategies. 
    }
   \label{fig:avglatIntegration}
\end{figure}

\textbf{Composite Metrics.} Fig. \ref{fig:avglatIntegration} confirms these findings using composite metrics. \framename achieves the lowest LCP (178.84 in Fig. \ref{subfig:lcp}) and IRI (683k in Fig. \ref{subfig:iri}). In contrast, baseline strategies show much higher inefficiency: \textit{Huawei} reaches an LCP of 478.64 and an IRI of 6169k, while \textit{DPSO} records an LCP of 217.96 and IRI exceeding 1,000k. These results demonstrate that \framename provides the most holistic balance between performance and sustainability.

\subsection{Performance on Challenging Workloads}
\label{subsec:taildistr}

We next evaluate \framename on the Long-tailed workload (64,375 invocations), a subset characterized by functions mainly with ``Custom" runtimes, high CPU and memory demand, and heavy initialization penalties, which are especially critical for latency-sensitive applications and sustainability-aware platforms.

\begin{figure}[t]
    \centering
    \subfloat[Cold Starts Count.]{\label{subfig:largelatcscount}\includegraphics[width=0.5\columnwidth]{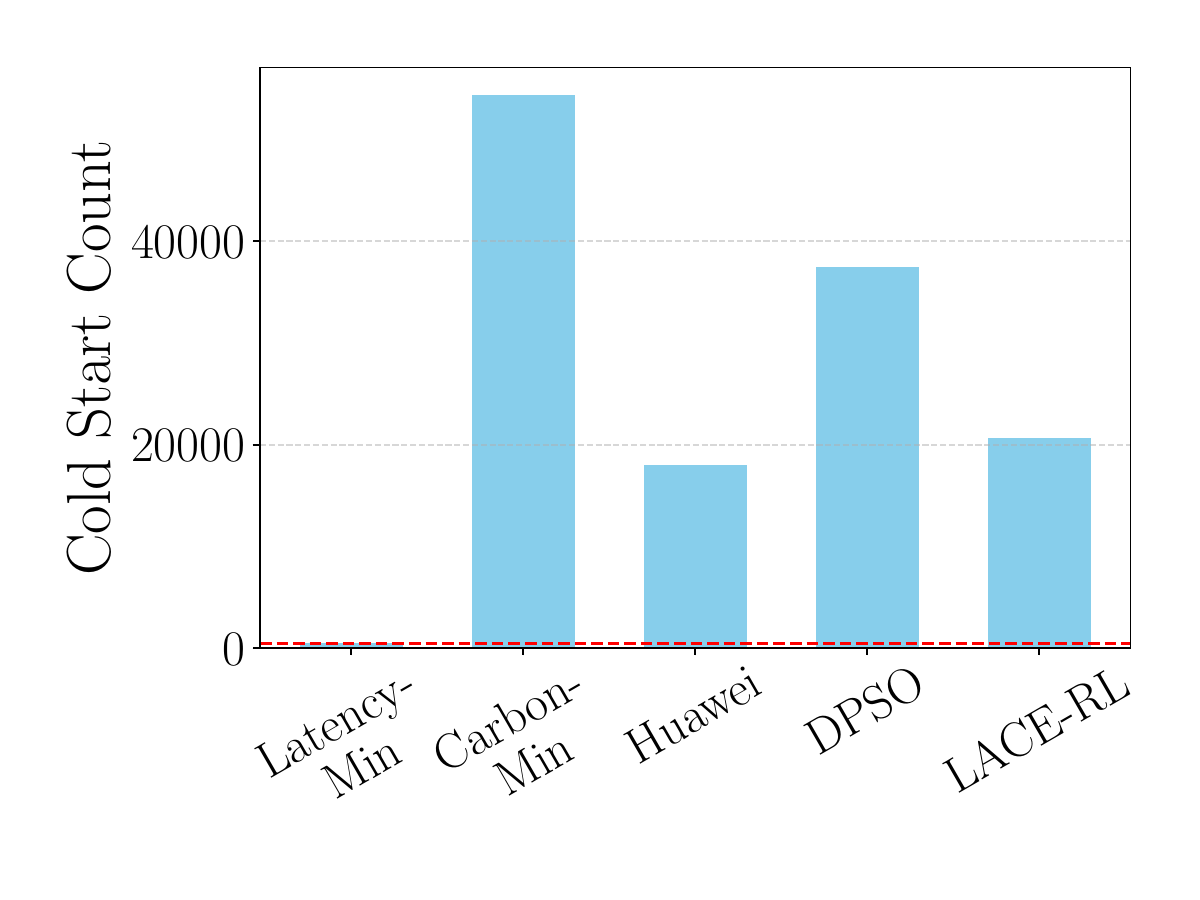}}
    \hfill
    \subfloat[End-to-end latency.]{\label{subfig:largelate2elatency}\includegraphics[width=0.5\columnwidth]{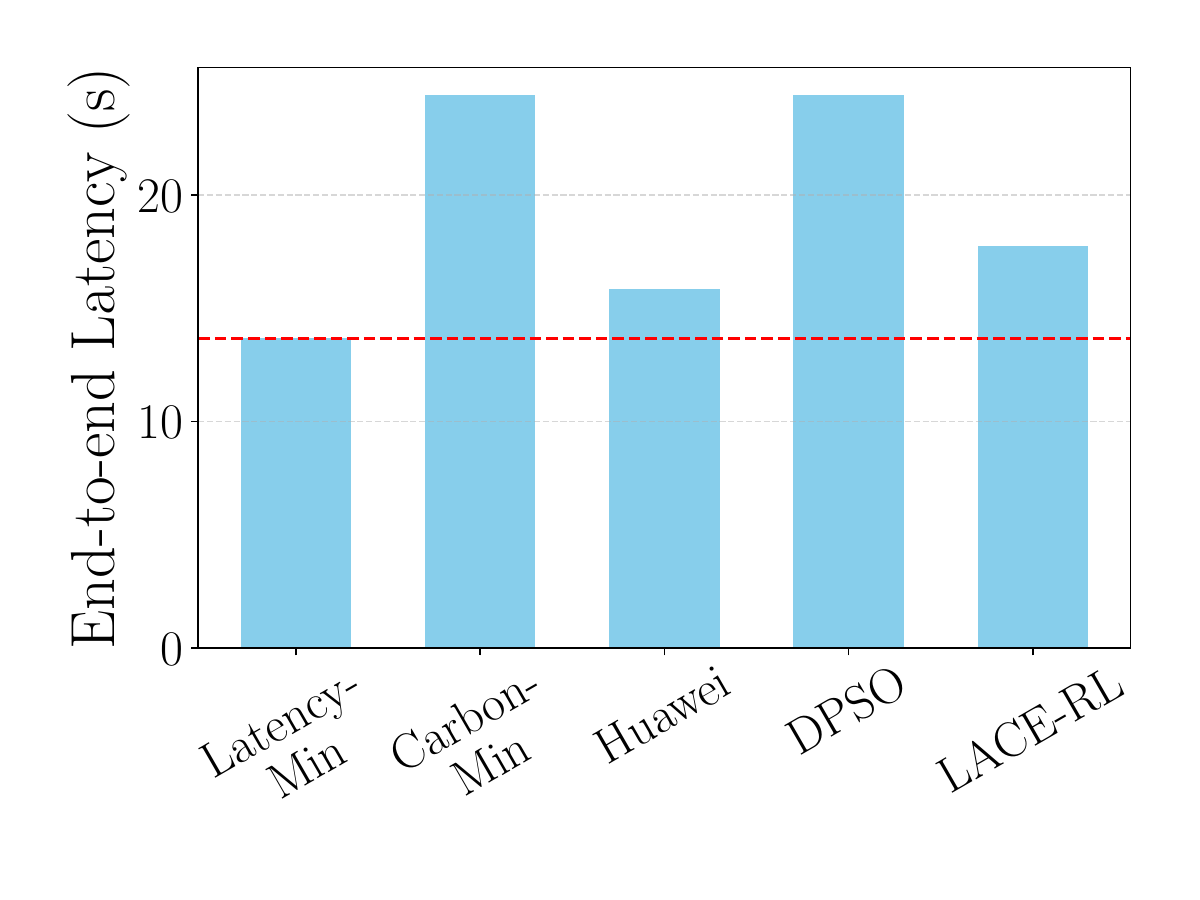}}
    \vfill
    \centering
    \subfloat[Keep Alive Carbon.]{\label{subfig:largelatkacarbon}\includegraphics[width=0.5\columnwidth]{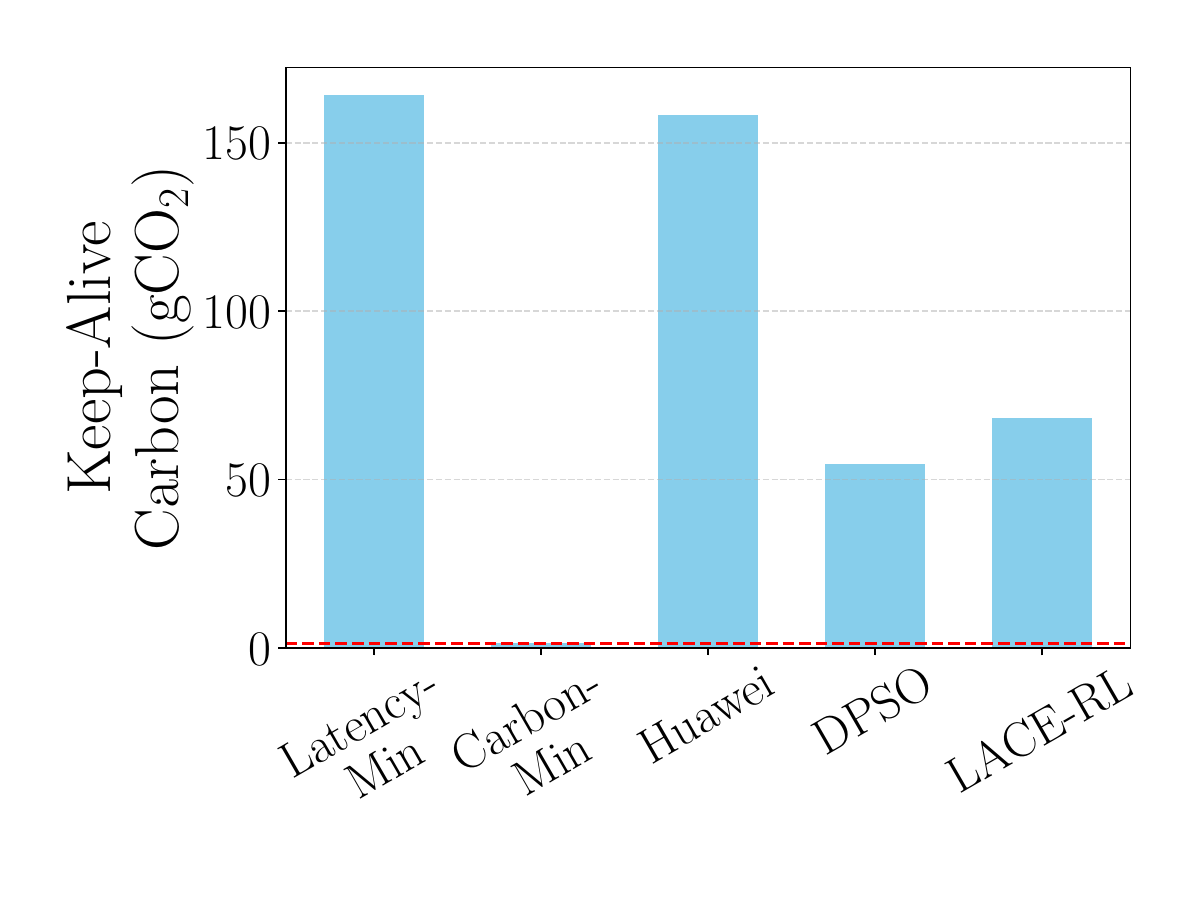}}
    \hfill
    \subfloat[Total Carbon Footprint.]{\label{subfig:largelattotcarbon}\includegraphics[width=0.5\columnwidth]{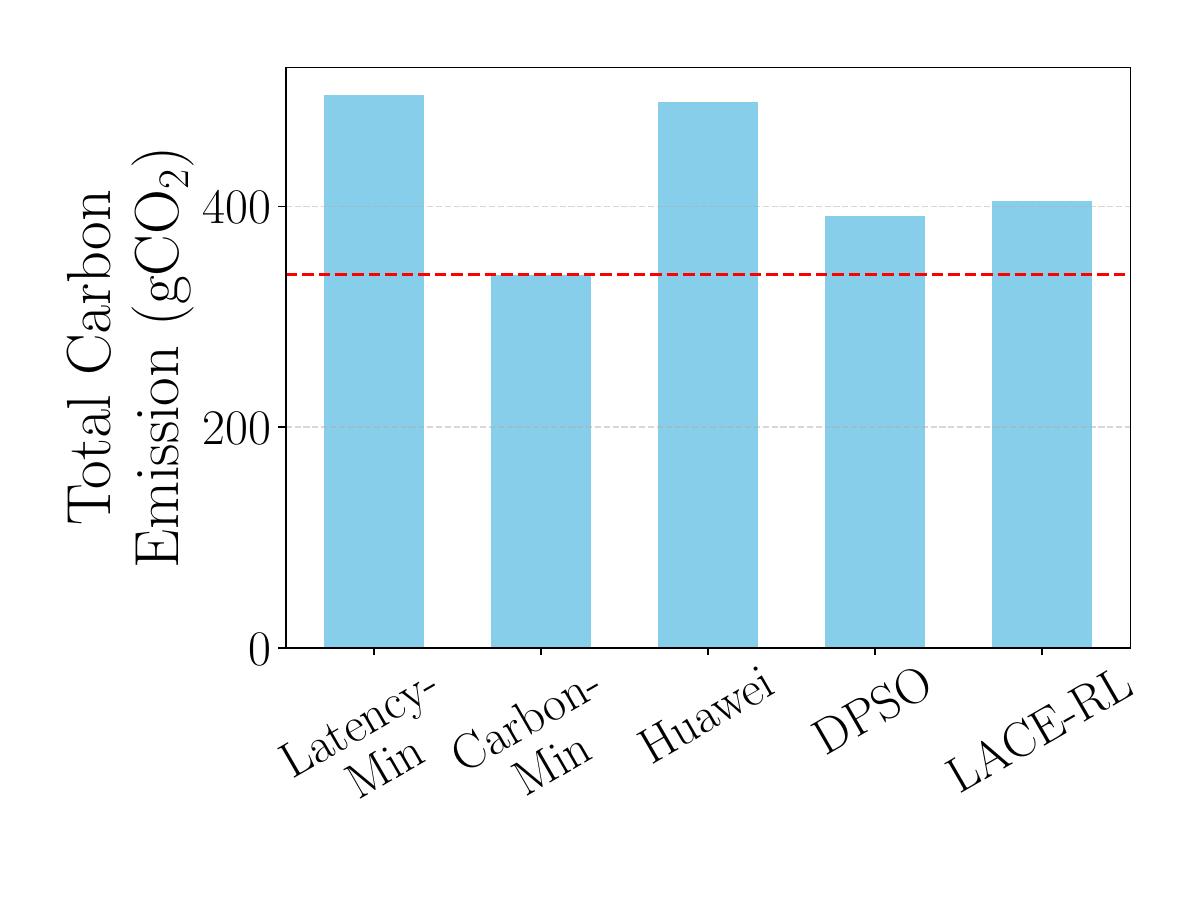}}
    \caption{Long-tailed workload results: the dashed line shows the optimal value.} 
    \label{fig:largelat}
\end{figure}

\textbf{Results.} As shown in Fig. \ref{fig:largelat}, \framename maintains good performance even in this demanding scenario.
\begin{itemize}[leftmargin=10pt]
    \item \textbf{Cold Starts:} \framename achieves 20,645 cold starts (Fig. \ref{subfig:largelatcscount}), fewer than \textit{Carbon-Min} (54,374) and \textit{DPSO} (37,485). While \textit{Latency-Min} minimizes counts (460), it triggers a massive surge in keep-alive carbon.
    \item \textbf{End-to-end Latency:} \framename achieves an average latency of 17.75s (Fig. \ref{subfig:largelate2elatency}), outperforming \textit{DPSO} (24.41s) and \textit{Carbon-Min} (24.41s), although slightly higher than \textit{Huawei} (15.86s) due to the trade-off with keep-alive carbon.
    \item \textbf{Carbon Emissions:} \framename emits only 68.15 gCO\textsubscript{2} for keep-alive (Fig. \ref{subfig:largelatkacarbon}). In contrast, \textit{Latency-Min} surges to 164.11 gCO\textsubscript{2} and \textit{Huawei} to 158.21 gCO\textsubscript{2}. Similar trend can be observed in the total carbon footprint in Fig. \ref{subfig:largelattotcarbon}. \framename achieves a total carbon footprint of 404.76 gCO\textsubscript{2}, lower than \textit{Huawei} and on par with \textit{DPSO}, but with lower latency.
\end{itemize}

\begin{figure}[t]
\centering\includegraphics[width=\columnwidth]{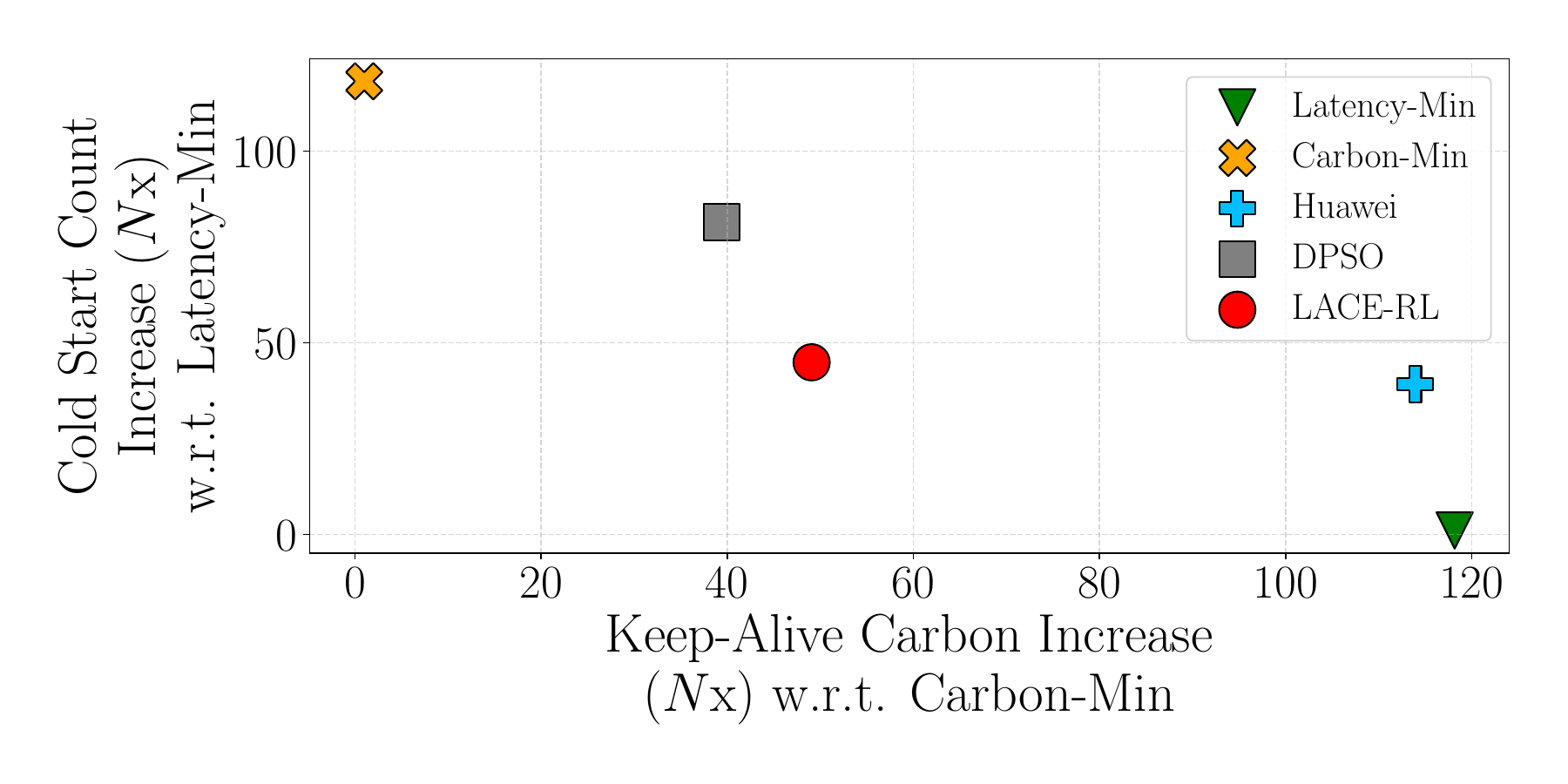}
    \caption{Long-tailed workload: \framename achieves the most balanced tradeoff between cold starts and carbon emissions.}
   \label{fig:largelatIntegration}
\end{figure}

\textbf{Tradeoff Visualization.}
Fig. \ref{fig:largelatIntegration} visualizes the normalized tradeoff. \framename identifies the optimal balance point, while \textit{Huawei} and \textit{DPSO} shift further from the origin, indicating reduced adaptability.
Composite metrics further validate this: \framename maintains the lowest LCP (7185.73) and IRI (1400k), surpassing \textit{Huawei} (LCP: 7845, IRI: 2050k) and \textit{DPSO} (LCP: 9553, IRI: 2850k). This confirms that \framename generalizes well to complex, high-latency functions.

\subsection{Comparing \framename with Oracle}
\label{subsec:oracle}
To quantify the gap between the prediction model of \framename and the optimum decision, we compare it against an Oracle policy (equipped with perfect future knowledge) over a two-hour trace slice. As shown in Table \ref{tab:oracle-comparison}, \framename incurs only a 6.18\% increase in keep-alive carbon and a 7.20\% increase in cold starts for the General workload.
For the Long-tailed workload, the gap widens slightly to 8.97\% (carbon) and 11.15\% (cold starts). This discrepancy is due to the bursty arrival patterns of long-tailed functions, which limit the effectiveness of window-based reuse prediction. However, \framename remains highly efficient, operating very close to the theoretical limit even under unpredictable conditions.

\begin{table}[t]
  \caption{Comparison of \framename vs. Oracle }
  \label{tab:oracle-comparison}
  \resizebox{\linewidth}{!}{
  \centering
  \begin{tabular}{c c c c c}
    \toprule
    Case & Metrics & Oracle & \framename & Degradation\\
    \midrule
    \multirow{2}{*}{General} & Keep-alive Carbon (gCO\textsubscript{2}) & 2.88 & 3.058 & 6.181\% \\
    & Cold Start Count (Times) & 1069 & 1146 & 7.203\% \\
    \midrule
    \multirow{2}{*}{Long-tailed} & Keep-alive Carbon (gCO\textsubscript{2}) & 3.9 & 4.25 & 8.974\% \\
    & Cold Start Count (Times) & 807 & 897 & 11.152\% \\
    \bottomrule
  \end{tabular}
  }
\end{table}

\subsection{Inference and Training Cost}
\label{subsec:inference}
Inference overhead is a critical factor for production deployment. We measure the inference time across the Long-tailed workload (64,375 invocations). The trained \framename agent completes inference for the entire workload in just 0.9642 seconds, averaging approximately 15 microseconds per invocation. 
While interactions with the underlying control plane (e.g., Kubernetes) to enforce these decisions may incur additional millisecond-level latency (e.g., via CRD updates), this overhead is asynchronous to the function execution path and negligible relative to the decision granularity. 
This low overhead makes it suitable for real-time, latency-sensitive environments.
In contrast, the heuristic-based \textit{DPSO} strategy requires 4,522.53 seconds for the same workload—over 4,600$\times$ slower than \framename. This is because DPSO performs iterative population updates for every decision, which scales poorly.

Regarding training, although the DQN requires offline training (approximately 5 minutes per episode), this cost is amortized over long-term deployment. 

\subsection{Sensitivity Analysis and Interpretability}
\label{subsec:sensitivity}
\begin{figure}[t]
    \centering
    \subfloat[Impact of $\lambda_{carbon}$ on cold start counts and keep-alive carbon.]{\label{subfig:lamcarbon}\includegraphics[width=0.48\columnwidth]{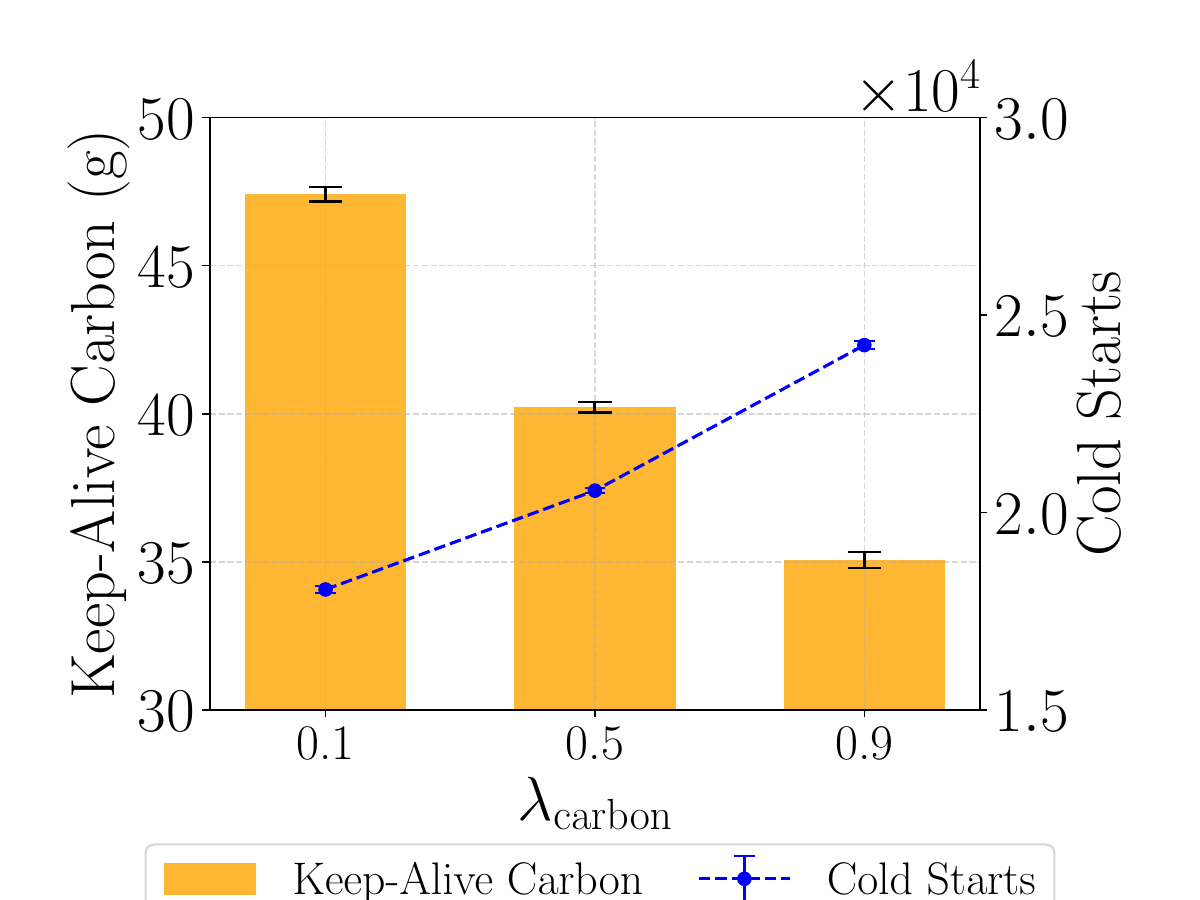}}
   \hspace{0.02\columnwidth}
   \subfloat[Percentage of three representative keep-alive durations by \framename as a function of hourly carbon intensity.]{\label{subfig:Freq}\includegraphics[width=0.48\columnwidth]{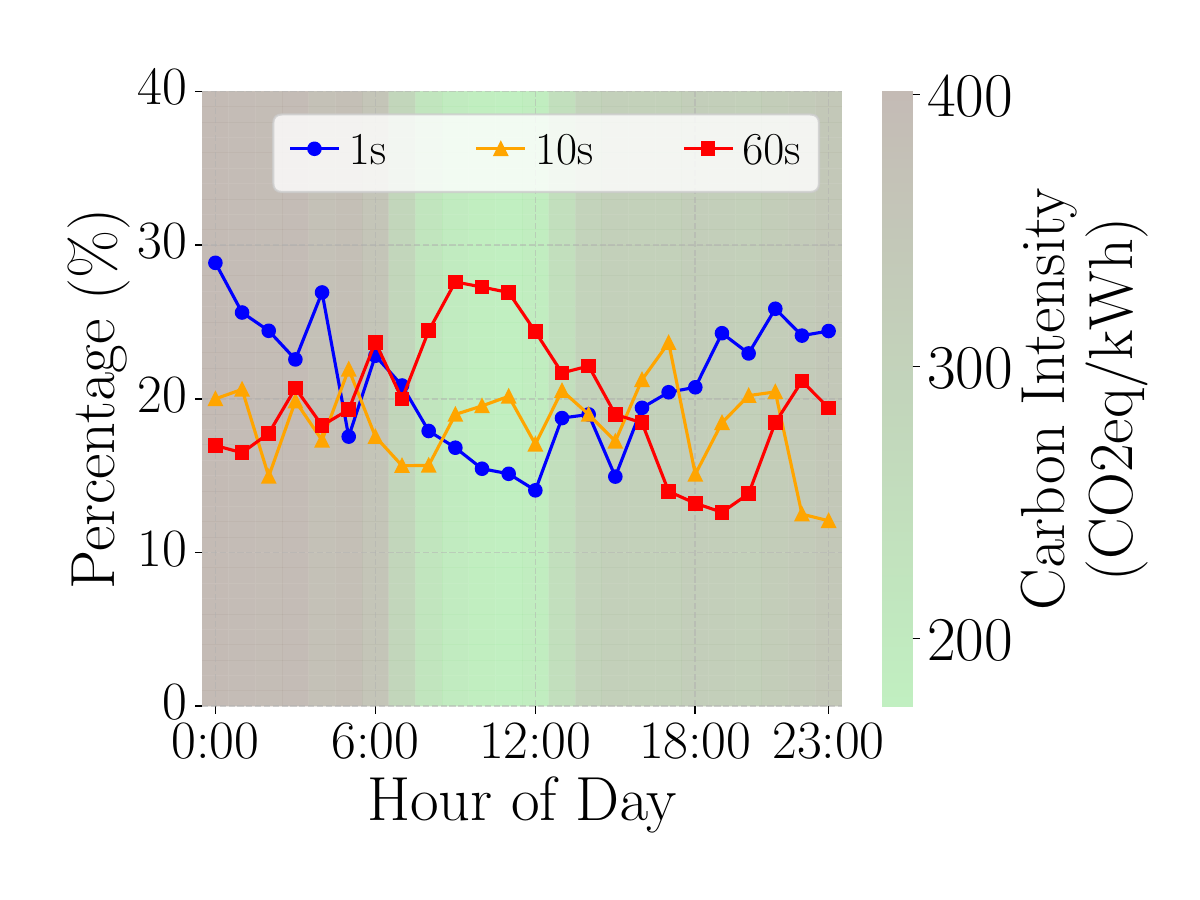}} 
    \caption{Sensitivity of \framename to tunable parameters (a) and interpretability (b).}
   \label{fig:lambda}
\end{figure}

We analyze how the tunable parameter $\lambda_{carbon}$ affects system behavior. Fig. \ref{subfig:lamcarbon} shows that as $\lambda_{carbon}$ increases from 0.1 to 0.9, the agent increasingly prioritizes carbon reduction, resulting in lower keep-alive emissions at the cost of higher cold start counts. This confirms that \framename provides stable, predictable control over the latency–carbon trade-off.

To interpret the learned strategy, Fig. \ref{subfig:Freq} plots the selection frequency of representative keep-alive durations against hourly carbon intensity.
During low-carbon hours (lighter background, e.g., 08:00–12:00), \framename frequently selects longer keep-alive durations (e.g., 60s) to minimize cold starts when energy is ``green". On the contrary, during high-carbon periods (darker background), the agent shifts toward shorter timeouts (e.g., 1s) to minimize idle emissions. This demonstrates that \framename successfully learns to adapt its strategy in response to real-time conditions of the electricity grid.

\section{Related Work}
\label{sec:related}

Optimizing performance and sustainability in serverless computing is of growing research interest~\cite{ebrahimi2024cold}. 
Prior efforts span several categories, including cold start mitigation~\cite{du2020catalyzer}, pre-warming strategies~\cite{hu2025mitigating}, carbon-aware management~\cite{gsteiger2024caribou, lin2024bridging}, and function invocation prediction~\cite{tomaras2023prediction, lee2024spes}. 
To our knowledge, \framename is the first RL-based framework that tunes keep-alive times to co-optimize cold-start latency and carbon emissions under time-varying grid carbon intensity.

\textbf{Cold Start Mitigation.}
Prior work has explored systems-level optimizations to reduce function startup time in FaaS platforms, such as Catalyzer~\cite{du2020catalyzer} and AWS Lambda’s Firecracker microVM technology~\cite{firecracker2025}.
While effective in reducing latency, these techniques prioritize performance only, often incurring memory or background service overhead~\cite{katsakioris2022faas}, and do not account for energy or carbon implications. In contrast, \framename incorporates the cold start mitigation.

\textbf{Pre-warming Techniques.}
Another class of solutions focuses on pre-warming or reusing pods across invocations. Many serverless platforms implement a fixed timeout as the keep-alive time~\cite{joosen2025serverless}, but do not adapt to workload diversity. 
Adaptive approaches adjust retention based on invocation frequency or classification~\cite{shahrad2020serverless, lee2024spes}. 
Vahidinia et al.\cite{vahidinia2022mitigating} combine reinforcement learning with LSTM models to reduce cold starts. 
These methods improve over static strategies but still rely on predefined rules and assume fixed cold start for all the functions.
\framename differs by dynamically adjusting keep-alive durations, based on reuse probability and cold start impact, enabling finer-grained and workload-aware strategies, and also incorporates carbon costs. Moreover, \framename models runtime-dependent cold-start latency rather than assuming a fixed cost.

\textbf{Carbon-Aware Management.} 
EcoLife~\cite{jiang2024ecolife} applies a PSO-based optimization framework to co-select keep-alive durations and hardware generations.
CarbonScaler~\cite{hanafy2023carbonscaler} dynamically scales serverless resources based on carbon intensity fluctuations in the grid, but targets delay-tolerant workloads and does not address interactive cold-start tradeoffs. Similarly, EcoFaaS~\cite{stojkovic2024ecofaas} redesigns serverless environments to reduce baseline energy waste, focusing on system-level efficiency rather than adaptive management. \framename differs from the above by integrating carbon-aware decision-making directly into an RL formulation that balances idle carbon and cold-start latency, allowing fine-grained, workload-aware decisions at runtime with microsecond-level inference overhead.

\section{Conclusion}
We introduce \framename, a reinforcement learning–based framework that adaptively tunes keep-alive durations by jointly modeling cold start probability, latency cost, and carbon emissions. \framename identifies the behavior of cold start latencies of different functions and adapts strategies accordingly.
Evaluation with the recently released Huawei trace shows that \framename substantially reduces both cold starts and carbon emissions while maintaining low latency, achieving performance close to an Oracle policy.

\section*{Acknowledgment}
This work has been supported in part by the Horizon Europe research and innovation programme of the European Union, under grant agreement no 101092912, project MLSysOps and the U.S.  National Science Foundation (NSF) grants \#2402942 and \#2403088.

\bibliographystyle{IEEEtran}
\bibliography{REF}

\end{document}